\documentclass[a4paper,11pt]{article}
\pdfoutput=1
\usepackage{jheppub} 
\usepackage{xcolor}

\title{\boldmath Lepton $g-2$ and $W$-boson mass anomalies\\
in the DFSZ axion model}

\author[a,b]{Moslem Ahmadvand,}
\author[c,d,e]{and Fazlollah Hajkarim}

\affiliation[a]{School of Particles and Accelerators, Institute Research in Fundamental Sciences (IPM)\\
P. O. Box 19395-5531, Tehran, Iran}
\affiliation[b]{School of Physics, Damghan University, Damghan 3671645667, Iran}
\affiliation[c]{Dipartimento di Fisica e Astronomia, Università degli Studi di Padova\\
Via Marzolo 8, 35131 Padova, Italy}
\affiliation[d]{Istituto Nazionale di Fisica Nucleare (INFN), Sezione di Padova\\
Via Marzolo 8, 35131 Padova, Italy}
\affiliation[e]{Homer L. Dodge Department of Physics and Astronomy, University of Oklahoma, \\
Norman, OK 73019, USA}

\emailAdd{ahmadvand@ipm.ir}
\emailAdd{fazlollah.hajkarim@pd.infn.it}

\abstract{With regard to the leptonic magnetic dipole moment anomaly as well as the $W$-boson mass excess, we study the DFSZ axion model.
Considering theoretical and experimental constraints, we show that the muon and electron $g-2$ anomalies can be explained within the parameter space of the model for extra Higgs bosons with mass spectra around the electroweak scale and for an almost equivalent contribution of one- and two-loop diagrams. A negative electron $g-2$ could be achieved by introducing heavy neutrinos. Furthermore, the $W$-boson mass excess can be consistently addressed within the mass range of the matter content testable at collider experiments. }


\begin{document} 
\begin{flushright}
\end{flushright}
\maketitle

\section{Introduction} \label{intro}




QCD axion models are well-motivated solutions to the strong CP problem of the Standard Model (SM) \cite{Peccei:1977hh, Peccei:1977np, Peccei:2006as}.
The axion appears as a pseudo-Nambu Goldstone boson from the spontaneous symmetry breaking of the $U(1)_\mathrm{PQ}$ Peccei-Quinn (PQ) symmetry at high temperatures and gains a mass at lower scales around the QCD phase transition~\cite{Weinberg:1977ma, Wilczek:1977pj}. The axion being neutral and long-lived, it is a viable candidate for cold dark matter. It could have been produced in the early universe by the misalignment mechanism or by decays of topological defects~\cite{Preskill:1982cy, Abbott:1982af, Dine:1982ah, Davis:1985pt}. A review of different axion models and their cosmological implications can be found in Refs.~\cite{Marsh:2015xka, DiLuzio:2020wdo}.

Although high-energy experiments have not reported so far any significant deviation from the SM, a set of experiments focusing on precision measurements may indicate anomalies in the SM requiring new physics. One of these high-precision experiments in particle physics measures the muon magnetic moment, $a_\mu \equiv (g_\mu-2)/2$, where $g_\mu$ is the spin $g$ factor of the muon. Fermi National Accelerator Laboratory (FNAL) has recently reported a large deviation from the SM prediction on $a_\mu $~\cite{Muong-2:2021ojo}, consistent with the previous measurement of Brookhaven National Laboratory (BNL) in 2006~\cite{Muong-2:2006rrc}. The result of combining these measurements is
\begin{equation} \label{delamu}
    \Delta a_\mu\equiv a_\mu^{\textrm{EXP}}-a_\mu^{\textrm{SM}}=(251\pm 59)\times 10^{-11},
\end{equation}
which shows a $4.2 \sigma$ discrepancy with the SM prediction~\cite{Aoyama:2020ynm, Aoyama:2012wk, Aoyama:2019ryr, Czarnecki:2002nt, Gnendiger:2013pva, Davier:2017zfy, Keshavarzi:2018mgv, Colangelo:2018mtw, Hoferichter:2019mqg, Davier:2019can, Keshavarzi:2019abf, Kurz:2014wya, Melnikov:2003xd, Masjuan:2017tvw, Colangelo:2017fiz, Hoferichter:2018kwz, Gerardin:2019vio, Bijnens:2019ghy, Colangelo:2019uex, Blum:2019ugy, Colangelo:2014qya}.

Moreover, improved measurements of the fine-structure constant $\alpha$ allow us to extract a value for the electron magnetic moment, $a_e$, which shows a discrepancy with the SM prediction. Interestingly, there are two experiments whose results on $a_e$ are set in opposite directions. Based on the recoil frequency of $^{133}$Cs atoms at Berkeley~\cite{Parker:2018vye}, one has
\begin{equation}\label{ber}
    \Delta a_e^{\textrm{B}}\equiv a_e^{\textrm{EXP}}-a_e^{\textrm{SM}}=(-87\pm 36)\times 10^{-14},
\end{equation} 
which features a $2.4\sigma$ discrepancy with the SM prediction~\cite{Hanneke:2008tm, Hanneke:2010au}, while with $^{87}$Rb atoms, the result of the Laboratoire Kastler Brossel (LKB)~\cite{Morel:2020dww} is
\begin{equation}\label{lkb}
    \Delta a_e^{\textrm{LBK}}\equiv a_e^{\textrm{EXP}}-a_e^{\textrm{SM}}=(48\pm 30)\times 10^{-14},
\end{equation} 
with a $1.6 \sigma$ deviation. We emphasize that even if both experiments show tension with respect to the SM in opposite directions, new physics may be required.

Additionally, another precise measured observable is the $W$-boson mass. Recently, the CDF-II experiment at Fermilab reported a new measurement of the $W$-boson mass, using data collected from 2002 to 2011~\cite{CDF:2022hxs}
\begin{equation}
    m_{W}^\mathrm{CDF-II}=80.4335\pm 0.0094~\mathrm{GeV},
\end{equation} 
which is in tension with the SM prediction $m_W^\mathrm{SM} = 80.357 \pm 0.006$~GeV~\cite{ParticleDataGroup:2022pth} about $7 \sigma$ and also differs from the previous average of PDG $m_{W}^\mathrm{PDG} = 80.379 \pm 0.0012$~GeV~\cite{ParticleDataGroup:2022pth}.  Once again, if such a deviation is confirmed, new physics is required.\footnote{Regarding uncertainties that may be originated from hadronic contributions, Ref.~\cite{Athron:2022qpo} demonstrates that the $W$-boson mass and $a_{\mu}$ anomalies pull the SM hadronic contributions in opposite directions and hence new physics is welcome.}

So far, a large number of different scenarios beyond the SM have been proposed to address the anomalies mentioned above (see, for example, Refs.~\cite{Popov:2016fzr, DelleRose:2020oaa, Cacciapaglia:2022xih, Lee:2022nqz, Babu:2022pdn, Balkin:2022glu, Ahn:2022xax, Kawamura:2022uft, Ghoshal:2022vzo, Kanemura:2022ahw, Chowdhury:2022moc, Borah:2022zim, Lee:2022gyf, Abouabid:2022lpg, Kim:2022hvh, Chowdhury:2022dps, Hessenberger:2022tcx, Saez:2021qta, Chakrabarty:2022gqi, Agrawal:2022wjm, Chakrabarty:2022voz, Abdallah:2022shy, Heckman:2022the, Hiller:2019mou,Han:2018znu}).
Here, alternatively, we employ the Dine-Fischler-Srednicki-Zhitnitsky (DFSZ) axion model~\cite{Zhitnitsky:1980tq, Dine:1981rt}, which is an attractive framework that not only addresses the strong CP problem but also provides a viable dark matter candidate.
The field content of the model, invariant under the global $U(1)_\mathrm{PQ}$, includes a Peccei-Quinn complex scalar field and one additional Higgs doublet.
We explore possible explanations for anomalous measurements and their phenomenological implications, restricting the parameter space of the model.
Considering vacuum stability and perturbativity conditions as well as collider bounds on the mass of extra Higgs bosons~\cite{ParticleDataGroup:2014cgo, CMS:2018qvj}, we show that for Higgs fields with masses at the electroweak (EW) scale, the CDF-II $W$-boson mass anomaly and the anomalous muon magnetic moment can be explained. Furthermore, the measurement of the anomalous electron magnetic moment can be fitted in the case of a positive contribution.
In the case of a negative $g_e-2$, taking into account the bounds on charged lepton flavor decay processes such as $\mu\to e\, \gamma$, extra degrees of freedom (d.o.f.) are required; for example, a heavy neutrino contribution can be added, without spoiling the solutions to the $W$ mass or $g_\mu-2$.

As we discussed, QCD axion is motivated as a solution to the strong CP problem and can be a candidate for dark matter. These are the low-energy phenomenological consequences of QCD axion models. 
Depending on the axion model, it can have different kinds of phenomenology at high energies. Here, our focus is on the DFSZ axion model to verify phenomenological effects imposed by possible explanations of the electron and muon magnetic moment and $W$-boson mass anomaly on the model.

In the next section, we introduce the model and discuss the physical d.o.f. and the mass spectrum. In Section~\ref{sec:wbm}, we discuss the oblique parameters, particularly the $\mathcal{T}$ parameter, and its impact on the $W$ mass.  Section~\ref{sec:gm2} is devoted to the one- and two-loop-level contributions to the lepton magnetic moment. We conclude in Section~\ref{sec:conc}.

\section{The Model} \label{model}
Regarding the smallness of the experimentally bounded CP-violating $\theta$ term in strong interactions, i.e. the strong CP problem, several QCD axion models, all based on a $U(1)_{\mathrm{PQ}}$ symmetry, have been proposed.
Different QCD axion models correspond to diverse realizations of the PQ symmetry under which the SM is not invariant.
In particular, to absorb the $\theta$ term, independent chiral transformations of the $u$ and $d$ quarks are required~\cite{Peccei:2006as}.
The model can therefore be enlarged with a second Higgs doublet coupled to only the $u$- or to the $d$-quark types via Yukawa interactions~\cite{Weinberg:1977ma, Wilczek:1977pj}.

Moreover, the PQ symmetry could be spontaneously broken at an energy scale much higher than the EW scale, leading to the generation of a pseudo-Goldstone boson, the axion. By imposing astrophysical constraints, the symmetry-breaking scale is bounded to $10^8~\mathrm{GeV}\lesssim f_a \lesssim 10^{17}$~GeV~\cite{Raffelt:2006cw, Arvanitaki:2009fg, DiLuzio:2020wdo}, where $f_a$ stands for the axion decay constant. This decoupling between the electroweak and PQ symmetry-breaking scales naturally occurs in the presence of a SM-singlet scalar field $\Phi$.
This kind of QCD axion model, which has such a field content, is called the DFSZ model.%
\footnote{An alternative approach to the invisible QCD axion is the Kim-Shifman-Vainshtein-Zakharov (KSVZ) model, where in addition to the PQ scalar  vector-like quarks are introduced~\cite{Kim:1979if, Shifman:1979if}.}

The Lagrangian of the model is given by
\begin{equation} \label{kin}
    \mathcal{L} \supset \frac12 \left|\partial_\mu\Phi\right|^2 + \left(D_\mu H_d\right)^{\dag} D^\mu H_d + \left(D_\mu H_u\right)^\dag D^{\mu} H_u - V(\Phi, H_u, H_d)\,,
\end{equation}
where $\Phi$ is the PQ scalar field, $H_u$ and $H_d$ denote the two Higgses, doublets under $SU(2)_L$.
The covariant derivative is defined as
\begin{equation}
    D_\mu H_{d,u} \equiv \partial_\mu H_{d,u} - i\, g\, \tau\cdot W_\mu\,  H_{d,u} -i\, g'\, Y\,  B_\mu\, H_{d,u} \,,
\end{equation}
where $W_\mu^a$ and $B_\mu $ are the $SU(2)_L$ and $U(1)_Y$ gauge fields, respectively, whose coupling constants are denoted by $g$ and $g'$, and $\tau_i$ correspond to the Pauli matrices.
The scalar potential is given by
\begin{align} \label{pot}	
    &V(\Phi, H_u, H_d) = \lambda_{\phi} \left(|\Phi|^{2} -V_{\phi}^{2}/2\right)^{2} + \left|H_{d}\right|^{2} \left(\kappa_{d}\, |\Phi|^{2} -\mu_{d}^{2}\right) + \left|H_{u}\right|^{2} \left(\kappa_{u}\, |\Phi|^{2} - \mu_{u}^{2}\right) \nonumber\\
    &\hspace{1.3cm} -\left(\kappa\, \Phi H_{u}^{\dagger}\, H_{d} + \text{H.c.}\right) + \lambda_{d} \left|H_{d}\right|^{4} + \lambda_{u} \left|H_{u}\right|^{4} + \lambda \left(\left|H_{u}\right|^{2}\left|H_{d}\right|^2-\left|H_{u}^{\dagger} H_{d}\right|^{2} \right) .
\end{align}

Finally, the Yukawa interactions are
\begin{equation} \label{yuk}
	y_u^{H_u}\, \overline{Q}_L\, \widetilde{H}_u\, u_R + y_d^{H_d}\, \overline{Q}_L\, H_d\, d_R + y_l^{H_d}\, \overline{\Psi}_{L_l}\, H_d\, l_R + \text{H.c.}
\end{equation}
where $\widetilde{H}_u \equiv -i\, \tau_2\, H^*_u$.
The model is invariant under the SM gauge symmetry group $SU(3)_C \otimes SU(2)_L \otimes U(1)_Y$ and under the $U(1)_{\mathrm{PQ}}$, and therefore the PQ charges of the fields in this model can be obtained accordingly~\cite{Ahmadvand:2021vxs}. In fact, from the $\kappa$ term of Eq.~\eqref{pot} and the orthogonality between PQ and corresponding hypercharge currents~\cite{DiLuzio:2020wdo}, we obtain PQ charges of scalar fields as $X_{H_u}=\cos^2\beta$, $X_{H_d}=-\sin^2\beta$, and $X_{\Phi}=1$. Moreover, supposing that left-handed fermions have no PQ charges, $X_{u_R}=X_{H_u}$, and $X_{d_R}=X_{l_R}=-X_{H_d}$.

The Higgs doublets can be expanded as~\cite{Espriu:2015mfa}
\begin{equation}
    H_{d} = \left(
                \begin{array}{l}
                    \alpha_{+} \\
                    \alpha_{0}
                \end{array}
            \right)
            \qquad \text{and}\qquad
    H_{u}=\left(
        \begin{array}{l}
            \beta_{+} \\
            \beta_{0}
        \end{array}\right),
\end{equation} 
and their vacuum expectation values (VEV) are given by $\langle\alpha_+\rangle = \langle\beta_+\rangle = 0$, $\langle\alpha_0\rangle = v_d$, $\langle\beta_0\rangle = v_u$, with $v^2 \equiv v_d^2 + v_u^2$, where $v \simeq 246$~GeV is the EW VEV.
Furthermore, $\langle\Phi\rangle = v_\phi \sim f_a$~\cite{DiLuzio:2020wdo}.  

Using the minimization of the potential, three parameters $\mu_d$, $\mu_u$, and $V_{\phi}$, can be fixed (see Appendix~\ref{min}); therefore, in the scalar sector, the free parameters are $\lambda_{\phi}$, $\tan\beta = t_\beta \equiv v_u/v_d$, $\kappa$, $\kappa_{d}$, $\kappa_{u}$, $\lambda_{u}$, $\lambda$, and $v_{\phi}$.
Finally, we note that $\lambda_d$ is fixed by imposing that one of the Higgs bosons has a mass of 125~GeV, as the SM Higgs.

\subsection{Physical degrees of freedom in the scalar sector}

In this section, we determine the mass spectrum of the scalar sector. From two complex doublets and a complex singlet, there are 10 d.o.f. Three combined Goldstone components are related to gauge bosons and are eliminated by gauge transformations, corresponding to longitudinal parts of the three massive gauge bosons~\cite{Branco:2011iw, Espriu:2015mfa}.
After rotating the interaction eigenstates (details of the transformation can be found in Appendix~\ref{rot}), the mass eigenstates can be identified.
The spectrum counts a pair of charged scalars $H^\pm$, a pseudoscalar $A$, three CP-even scalars $h_i$, and the axion field $a$.

First, the mass of the pseudoscalar Higgs $A$ is given by 
\begin{equation}
	m_{A}^{2} = 8\, \frac{\kappa}{v_{\phi}}\left(s_{2 \beta}\, v^{2}+\frac{v_{\phi}^{2}}{s_{2 \beta}}\right).
\end{equation}
From Eq.~\eqref{pot}, the mass of the charged Higgs fields $H^\pm$ is
\begin{equation}
	m_{H^{\pm}}^{2}=8\left(\lambda\, v^{2} + \frac{\kappa\, v_{\phi}}{s_{2 \beta}}\right).
\end{equation}
Furthermore, there are three neutral scalar states, $H$, $S$, and $\varrho$, which are not mass eigenstates and should be expressed in the mass basis where the mass matrix can be diagonalized by a rotation matrix (Appendix~\ref{diag})
\begin{equation}\label{massmat}
	H=\sum_{i=1}^{3} R_{H i} h_{i}, \quad S=\sum_{i=1}^{3} R_{S i} h_{i}, \quad \varrho=\sum_{i=1}^{3} R_{\rho i} h_{i} \,.
\end{equation}
where $h_i$ are mass eigenstates and the rotation matrix, $R$, and its components are expressed in Appendix~\ref{diag}.
The mass spectrum of these scalar fields, up to $\mathcal{O}(v^2/v_{\phi}^2)$, is
\begin{align} \label{mspec}
    m_{h_{1}}^{2} &\simeq 32\, v^{2} \left(\lambda_{d}\, c_{\beta}^{4} + \lambda_{u}\, s_{\beta}^{4}\right) - \frac{16\, v^{2}}{\lambda_{\phi}} \left(\kappa_{d}\, c_{\beta}^{2} + \kappa_{u}\, s_{\beta}^{2} - \frac{\kappa}{v_{\phi}}\, s_{2 \beta}\right)^2, \nonumber\\
    m_{h_{2}}^{2} &\simeq \frac{8\, \kappa}{s_{2 \beta}} v_{\phi} + 8\, v^{2}\, s_{2 \beta}^{2}\left(\lambda_{d} + \lambda_{u}\right) - 4\, v^{2}\, \frac{\left[(\kappa_{d} - \kappa_{u})\, s_{2 \beta} + 2\, \kappa\, c_{2 \beta} / v_{\phi}\right]^{2}}{\lambda_{\phi} - 2\, \kappa/(v_{\phi}\, s_{2 \beta})} \,,\nonumber\\
    m_{h_{3}}^{2} &\simeq 4 \lambda_{\phi} v_{\phi}^{2} + 4 v^{2} \frac{\left[(\kappa_{d}-\kappa_{u}) s_{2 \beta}+2 \kappa c_{2 \beta}/v_{\phi}\right]^{2}}{\lambda_{\phi}-2 \kappa/(v_{\phi} s_{2 \beta})} + \frac{16 v^{2}}{\lambda_{\phi}} \left(\kappa_{d} c_{\beta}^{2}+\kappa_{u} s_{\beta}^{2}-\kappa s_{2 \beta}/v_{\phi}\right)^{2}.
\end{align}
The lighter state $h_1$ is identified with the SM-like Higgs boson, while $h_2$ and $h_3$ are extra heavy neutral CP-even Higgs bosons.

We do not consider cases where the mass spectrum is decoupled from the EW energy scale,  but cases that can address the aforementioned anomalous problems and lead to rather light masses for the Higgs fields,  phenomenologically interesting for collider physics.
In this sense, after fixing $t_{\beta}$, the mass spectrum of the Higgs fields depends mainly on $\kappa$, $\lambda$, and $\lambda_{\phi}$.
We consider $m_{h_1}$ as the SM Higgs and also fix $\lambda_{\phi} = 0.01$ that affects $m_{h_3}$ so that the very massive scalar field can be determined close to $v_\phi$ due to $v_\phi \gg v$.\footnote{We consider $\lambda_{\phi} = 0.01$ as a representative value and lighter $h_3$ can also be obtained for other allowed smaller values of $\lambda_{\phi}$.} The case with $\lambda_\phi \to 0 $ can lead to masses lighter than the SM Higgs mass.
SM-like interactions of the Higgs, $h_1$, with SM particles can also consistently be obtained, due to the relevant coefficient, a combination of $R_{H 1}\sim 1 $ and $R_{S 1}$ proportional to $ v^2/v_{\phi}^2$, multiplied to SM Higgs couplings. For more details see Ref.~\cite{Espriu:2015mfa}.

\section{\boldmath Collider constraints} \label{sec:cc}
Constraints on the mass spectrum of extra Higgs fields in the framework of the Minimal Supersymmetric Standard Model (MSSM)~\cite{MSSMWorkingGroup:1998fiq} and two-Higgs-doublet models (2HDMs)~\cite{Branco:2011iw}  are extensively studied~\cite{ParticleDataGroup:2022pth}. Regarding the mass of the charged Higgs boson, very light masses, $m_{H^{\pm}}\lesssim 80\,\mathrm{GeV}$ have been excluded by LEP~\cite{ALEPH:2013htx} and Tevatron~\cite{CDF:2009efz}. Moreover, for $m_{H^{\pm}}<m_t$ (the top quark mass), the experimental lower bound from direct searches is $m_{H^{\pm}}>155\,\mathrm{GeV}$~\cite{ParticleDataGroup:2022pth}. For mass ranges $m_{H^{\pm}}\sim 155-170\,\mathrm{GeV}$, firm experimental analysis does not exist~\cite{LHCHiggsCrossSectionWorkingGroup:2016ypw} and a reliable perturbative calculation of the charged Higgs boson production cross section must be performed. For $m_{H^{\pm}}>m_t$, ATLAS and CMS have also been excluded regions of the parameter space of MSSM, $m_{H^{\pm}}< 180 (1100)\,\mathrm{GeV}$, for $t_{\beta}=10 (60)$, with a certain luminosity~\cite{ParticleDataGroup:2022pth}. This search is sensitive to the modeling of the top pair production background with extra partons and especially b-quarks.

Also, there have been numerous experimental searches at the LHC for a pseudo-scalar boson, excluding light pseudoscalar Higgs bosons $m_{A}\lesssim m_{h_1}/2$~\cite{CMS:2018qvj}. In addition, in the context of MSSM, for heavy neutral/pseudo-scalar Higgs bosons in the $\tau\tau$ final state, the exclusion region is $m_A\lesssim 390 (1600)\,\mathrm{GeV}$ for $t_{\beta}=10 (60)$~\cite{ParticleDataGroup:2022pth}. However, in the diphoton channel a slight excess has been observed by CMS at a mass of $95.3\,\mathrm{GeV}$, consistent with the observed excess at LEP, with a local significance of $ 2.8\sigma$ and could not be ruled out by ATLAS at $95\% $ CL~\cite{ParticleDataGroup:2022pth}.

Experimental searches including extra Higgs production crucially depend on accurate theoretical predictions for inclusive cross sections. (For instance see Ref.~\cite{Ahmed:2016otz} for precise predictions of pseudo-scalar inclusive cross section specially for the moderate mass range.) According to the matter content of our model and its parameter space, given additional decay channels, for instance $H^{\pm}\rightarrow aW^{\pm} $, $A\rightarrow a h_1 $, $h_2\rightarrow aa $, and $h_2\rightarrow aZ $, the bounds may somehow change.\footnote{We leave precise cross section calculations to a future work.} Here, aside from the exclusion region independent from $t_{\beta}$, i.e., $m_{H^{\pm}}\lesssim 155\,\mathrm{GeV}$ and $m_{A}\lesssim m_{h_1}/2$, we consider other mass values, provided that the model parameter $t_{\beta}$ is in the viable, allowed range. In the following calculations, we show that the anomalies may simultaneously be explained for $t_{\beta}>100$.

\section{\boldmath $W$-boson mass} \label{sec:wbm}
In this section, we study the effects of new physics on the recently reported $W$-boson mass excess by analyzing the oblique parameters of this model~\cite{Peskin:1990zt, Peskin:1991sw}.
Some studies have already considered the impact of having a second Higgs doublet in the $W$-mass anomaly; see, e.g. Refs.~\cite{Babu:2022pdn, Kim:2022hvh, DelleRose:2020oaa, Ghoshal:2022vzo, Kanemura:2022ahw, Popov:2016fzr, Abouabid:2022lpg, Lee:2022gyf, Ahn:2022xax, Hessenberger:2022tcx, Borah:2022zim, Chowdhury:2022dps, Ghorbani:2022vtv, Chakrabarty:2022gqi}.
The leading correction to the $W$-boson mass can be approximated as
\begin{equation}
	\Delta m_W^2 \simeq \alpha\, \frac{c_W^4}{c_W^2-s_W^2}\, m_Z^2\, \Delta\mathcal{T}\,,
\end{equation}
where $\Delta\mathcal{T}$ is one of the Peskin-Takeuchi parameters, obtained by the following procedure. 
The one-loop corrected masses of the gauge bosons are given by
\begin{align}
	m_{W}^{2} &\simeq \frac{g^{2}\, v^{2}}{2}+\Pi_{W W}(0)\,, \\
	m_{Z}^{2} &\simeq \left(g^{2} + {g'}^2\right) \frac{v^{2}}{2}+\Pi_{Z Z}(0)\,,
\end{align}
where $\Pi_{WW}(0)$ and $\Pi_{ZZ}(0)$ are the two-point functions of the gauge fields at zero-momentum transfer~\cite{Espriu:2015mfa}.
The deviation from the custodial symmetry can be parameterized as
\begin{equation}
	\rho \equiv \frac{m_{W}^{2}}{m_{Z}^{2}\, c_{W}^{2}} = \frac{m_{W(\text {tree})}^{2} + \Pi_{W W}(0)}{m_{Z(\text {tree})}^{2}\, c_{W}^{2} + \Pi_{Z Z}(0)\, c_{W}^{2}} \simeq 1 + \frac{\Pi_{W W}(0)}{m_{W(\text {tree})}^{2}} - \frac{\Pi_{ZZ}(0)}{m_{Z(\text {tree})}^{2}}\,,
\end{equation}
where $m_{V (\text {tree})}^{2}$ is the mass at tree level of the gauge boson $V$, and 
\begin{equation}
	\Delta\rho \equiv \frac{\Pi_{W W}(0)}{m_{W(\text {tree})}^{2}} - \frac{\Pi_{ZZ}(0)}{m_{Z(\text {tree})}^{2}}\,.
\end{equation}
Thus, $\Delta\mathcal{T} = \Delta\rho/\alpha$ is~\cite{Espriu:2015mfa}
\begin{align}
    \Delta\mathcal{T} &= \frac{1}{16 \pi s_W^2m_W^2}\left[m_{H^{\pm}}^{2}-\frac{v_{\phi}^{2}}{v_{\phi}^{2}+v^{2} s_{2 \beta}^{2}} f\left(m_{H^{\pm}}^{2}, m_A^2\right)\right.\nonumber\\
    &\qquad \left.+\sum_{i=1}^{3} R_{S i}^{2}\left(\frac{v_{\phi}^{2}}{v_{\phi}^{2}+v^{2} s_{2 \beta}^{2}} f\left(m_A^{2}, m_{h_{i}}^{2}\right)-f\left(m_{H^{\pm}}^{2}, m_{h_{i}}^{2}\right)\right)\right] \nonumber\\
    &\simeq\frac{1}{16 \pi s_W^2m_W^2}\left[m_{H^{\pm}}^{2}- f\left(m_{H^{\pm}}^{2}, m_A^2\right)+\sum_{i=1}^{3} R_{S i}^{2}\left( f\left(m_A^{2}, m_{h_{i}}^{2}\right)-f\left(m_{H^{\pm}}^{2}, m_{h_{i}}^{2}\right)\right)\right],
\end{align}
where $R_{Si}$ is given in Appendix~\ref{diag}, $f(x, y) \equiv x\, y\, \ln (x/y)/(x-y) $ and $f(x, x) = x$. 
As a result, a positive value of $\Delta\mathcal{T}$ can explain the $W$-boson mass excess. We calculate the $W$-boson mass excess as a function of $m_{H^{\pm}}$, fixing all other parameters and masses. For a given value of $m_{A} $ and $m_{h_2} $, the excess, including the CDF-II measurement, can be justified within $ m_{H^{\pm}}/m_{A}\gtrsim 1$ or in some regions for $ m_{H^{\pm}}/m_{A}\lesssim 1$, except for $m_{H^{\pm}}=m_{A}$, $ m_{h_2} $ where $\Delta\mathcal{T}=0 $. Note that within the considered parameter space the $h_3$ contribution is negligible. In this case, the low-energy effective theory would be a 2HDM model that has the advantage of avoiding the strong CP problem thanks to the QCD axion. To relate the parameters of the DFSZ potential to those of a 2HDM, see Ref~\cite{Espriu:2015mfa}. Also, in the whole studied problems, we also subtract the SM-like Higgs contribution in the calculation.

\begin{figure}
	\centering
	\includegraphics[scale=1]{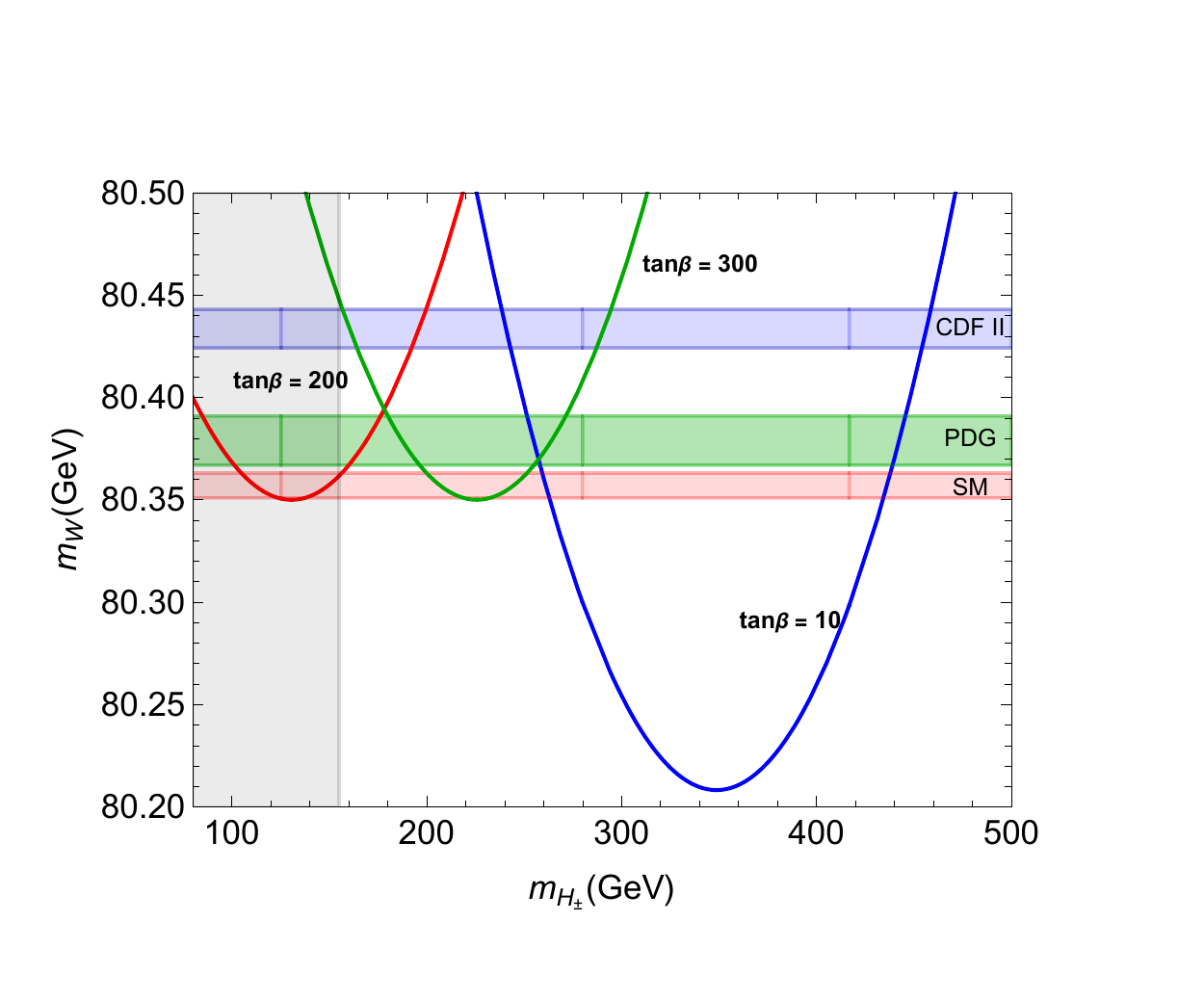}
        \vspace{0cm}
	\caption{
	The $W$-boson mass $m_{W}$ is shown as a function of $m_{H^{\pm}}$, for $\kappa_{d} = \kappa_{u} = v^2 /v_\phi^2$, $m_{h_1}=125$~GeV, $\lambda_{d}-\lambda_{u}=5$,  $\lambda_{\phi}=0.01$, $v_{\phi}=10^9$~GeV and three different values of $t_{\beta}=10, 200, 300$ and $\tilde{\kappa}=0.032, 0.00035, 0.001$ respectively. The relevant $W$-boson mass excess explanation with larger $m_{H^{\pm}}$ can be found with larger $\tilde{\kappa}$.
    } 
	\label{w mass}
\end{figure}

As we shall discuss in the following sections, the region of the parameter space that fits the lepton $g_l-2$ anomalies corresponds to small values of $\kappa$, $\kappa_{d}$ and $\kappa_{u}$, therefore, we fix $\kappa_{d} = \kappa_{u} = v^2 /v_\phi^2$ and $\kappa = \tilde{\kappa}\, v^2 / v_\phi$ with $\tilde{\kappa} < 1$.\footnote{Such small values of couplings are also consistent with the notion of naturalness and enhancing the Poincar\'e symmetry~\cite{Foot:2013hna}. }
Furthermore, considering vacuum stability conditions (Appendix~\ref{min}), we impose  perturbativity criteria as $\lambda,\, \lambda_{d,u}\lesssim 4\pi$.
Therefore, in Fig.~\ref{w mass}, we set other parameters as follows. Taking into account $m_{h_1}=125$~GeV, we can find $\lambda_{d,u} $, fixing $t_{\beta}$ and $\lambda_{d}-\lambda_{u}$. In this sense, since we obtain $\lambda_{d}-\lambda_{u}\approx \lambda_{d}$, we can fix $\lambda_{d}-\lambda_{u}< 4\pi$. In our calculations, we take $\lambda_{d}-\lambda_{u}=5$. Within the considered parameter values, we can approximately express the mass terms of scalar fields, Eq.\,(\ref{mspec}), with their first term for each relation.

Additionally, based on the perturbativity constraints from the Yukawa sector
Eq.~\eqref{yuk}, we consider a conservative upper bound $t_{\beta}\leq 300$. According to the mass relation of the scalar fields, $m_{h_3}$, $m_{h_2}$, and $m_{A}$ are also determined by fixing $\lambda_{\phi}$, and then $\kappa$. Thus, we can plot $m_{W}$ as a function of $m_{H^{\pm}}$. We consider $m_{H^{\pm}}$ as a variable, although its different values can be obtained by fixing $\lambda$ and $\tilde{\kappa}$. As shown in Fig.~\ref{w mass}, we can obtain the $W$-boson mass excess within the parameter space of the theory, showing by three different benchmarks for $\kappa_{d} = \kappa_{u} = v^2 /v_\phi^2$, $m_{h_1}=125$~GeV, $\lambda_{d}-\lambda_{u}=5$,  $\lambda_{\phi}=0.01$, $v_{\phi}=10^9$~GeV and three different values of $t_{\beta}=10, 200, 300$ and $\tilde{\kappa}=0.032, 0.00035, 0.0007$ respectively, which are corresponded to $\{m_{A}=280$~GeV, $m_{h_2}=417$~GeV, $m_{h_3}=0.2 v_{\phi}\}$, $\{m_{A}=130$~GeV, $m_{h_2}=131$~GeV, $m_{h_3}=0.2 v_{\phi}\}$, and $\{m_{A}=226$~GeV, $m_{h_2}=226$~GeV, $m_{h_3}=0.2 v_{\phi}\}$.

\section{\boldmath Lepton $g_l - 2$ anomalies} \label{sec:gm2}
In this section, we first compute and discuss the analytical expressions for the lepton $g_l - 2$ anomalies, and then we proceed to calculate the magnetic moment of the muon and electron.
On the basis of Lorentz covariance and gauge symmetry, the vertex of the QED interaction of a lepton can be expressed as
\begin{equation}
    \bar{l}\left(p'\right)\, \Gamma^\mu\, l\left(p\right) = \bar{l}\left(p'\right) \left[\gamma^\mu\, F_1\left(q^2\right) + \frac{i\, \sigma^{\mu \nu}\, q_\nu\, F_2\left(q^2\right)}{2\, m_l}\right] l\left(p\right),
\end{equation}
where $F_1(q^2)$ is related to the electric charge, and the lepton magnetic moment is defined as $a_l = F_2(0)$. Here, we calculate different $\Delta a_l$ contributions from the model.

Regarding the Yukawa couplings, using a unitary transformation on the charged leptons, these Yukawa coupling matrices can be diagonalized, $\tilde y_l $, so that from Eqs.~\eqref{yuk} and~\eqref{transh}, the Yukawa interactions of the leptonic sector, contributing to $a_{l}$, at one-loop level, can be expressed as 
\begin{equation}\label{yus}
	-\mathcal{L}_{Y}^{l}\supset -\frac{m_{l}}{v}\left( H\, \bar{l}\, l - t_{\beta}\, S\, \bar{l}\, l - i\, t_{\beta}\, \tilde{A}\, \bar{l}\, \gamma_{5}\, l\right).
\end{equation} 
 
\begin{figure}
	\centering
	\includegraphics[scale=1.4]{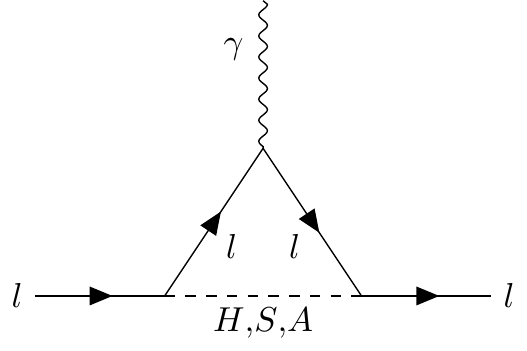}
	\caption{One-loop Feynman diagrams contributing to $g_l - 2$ from Higgs fields.}
	\label{feyn1}
\end{figure}

Therefore, from Eqs.~\eqref{massa} and~\eqref{massmat} for mass eigenstates we obtain the one-loop amplitude, Fig.~\ref{feyn1}, as follows (checked by using {\tt Package-X}~\cite{Patel:2015tea})
\begin{equation} \label{1lp}
    \Delta a_{l}^{1L} =\frac{m_{l}^2}{16 \pi^{2}\, v^2} \left[\sum_{i=1}^3 \left(R^2_{H_i} + t_{\beta}^{2}\, R^2_{S_i}\right)\, F(x_{h_i}) + \frac{t_{\beta}^{2}\, v_{\phi}^{2}}{v_{\phi}^{2} + v^{2}\, s_{2 \beta}^{2}}\, F(x_{A})\right],
\end{equation}
where $x_{h_i}=m_{l}/m_{h_i}$, and $x_{A}=m_{l}/m_{A}$.
\footnote{The contribution of the axion to Eq.~\eqref{1lp} is proportional to $\frac{t_{\beta}^{2}\, v^{2}\, s_{2 \beta}^{2}}{v_{\phi}^{2} + v^{2}\, s_{2 \beta}^{2}}\, F(x_{a})$ and is therefore negligible. For the contribution of EW gauge bosons, see, e.g. Ref.~\cite{Keshavarzi:2021eqa}.}  The loop functions obtained, $F(x)$ and $F_{H^{\pm}}(x, z)$ are given in Appendix~\ref{app-loop}.  

\begin{figure}
	\centering
	\includegraphics[scale=1]{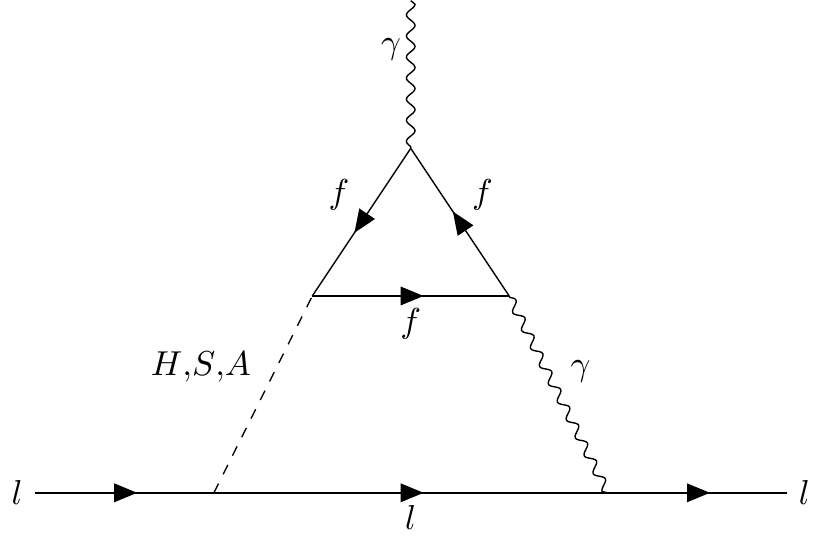}
	\caption{The Barr-Zee two-loop diagram contributing to $g_l - 2$.}
	\label{feyn2}
\end{figure}

According to Eq.~\eqref{yuk}, the so-called Barr-Zee two-loop amplitudes can be obtained~\cite{Barr:1990vd, Ilisie:2015tra, Cherchiglia:2016eui}, where the dominant contribution, Fig.~\ref{feyn2}, is given by 
\begin{equation}
    \Delta a_{l}^{2L} = \sum_{f} \frac{\alpha\, m_l^2}{4 \pi^3\, v^2}\, N_C^f\, Q_f^2 \left[\sum_{i=1}^3 \left(R^2_{H_i}\, \zeta_{f,l}^H + R^2_{S_i}\, \zeta_{f,l}^S\right) \mathcal{F}\left(\omega_{h_i}\right) + \frac{\zeta_f^A\, v_\phi^2}{v_\phi^2 + v^2\, s_{2 \beta}^2} \tilde{\mathcal{F}}\left(\omega_A\right)\right],
\end{equation}
where $N_C^f$ is the number of colors and $Q_f$ is the electric charge. Note that the two-loop contribution is at the $\alpha$ and $\tilde{y}_l^2$ level. 
We consider the dominant contribution from the $t$ and $b$ quarks, and therefore $\zeta_{t,l}^H=\zeta_{b,l}^H=1$, $\zeta_{t,l}^{S,A}=-1$, $\zeta_{b,l}^{S,A}=t_{\beta}^{2}$. Also, 
$\omega_{h_i}=m_f^2/m_{h_i}^2$, $\omega_{h_i}=m_f^2/m_{A}^2$. The functions $\mathcal{F}(\omega)$ and $\tilde{\mathcal{F}}(\omega)$ are defined in Appendix~\ref{app-loop}.

\subsection[Muon $g_\mu - 2$ anomaly]{\boldmath Muon $g_\mu - 2$ anomaly}
\begin{figure}
    \centering
    \includegraphics[scale=1]{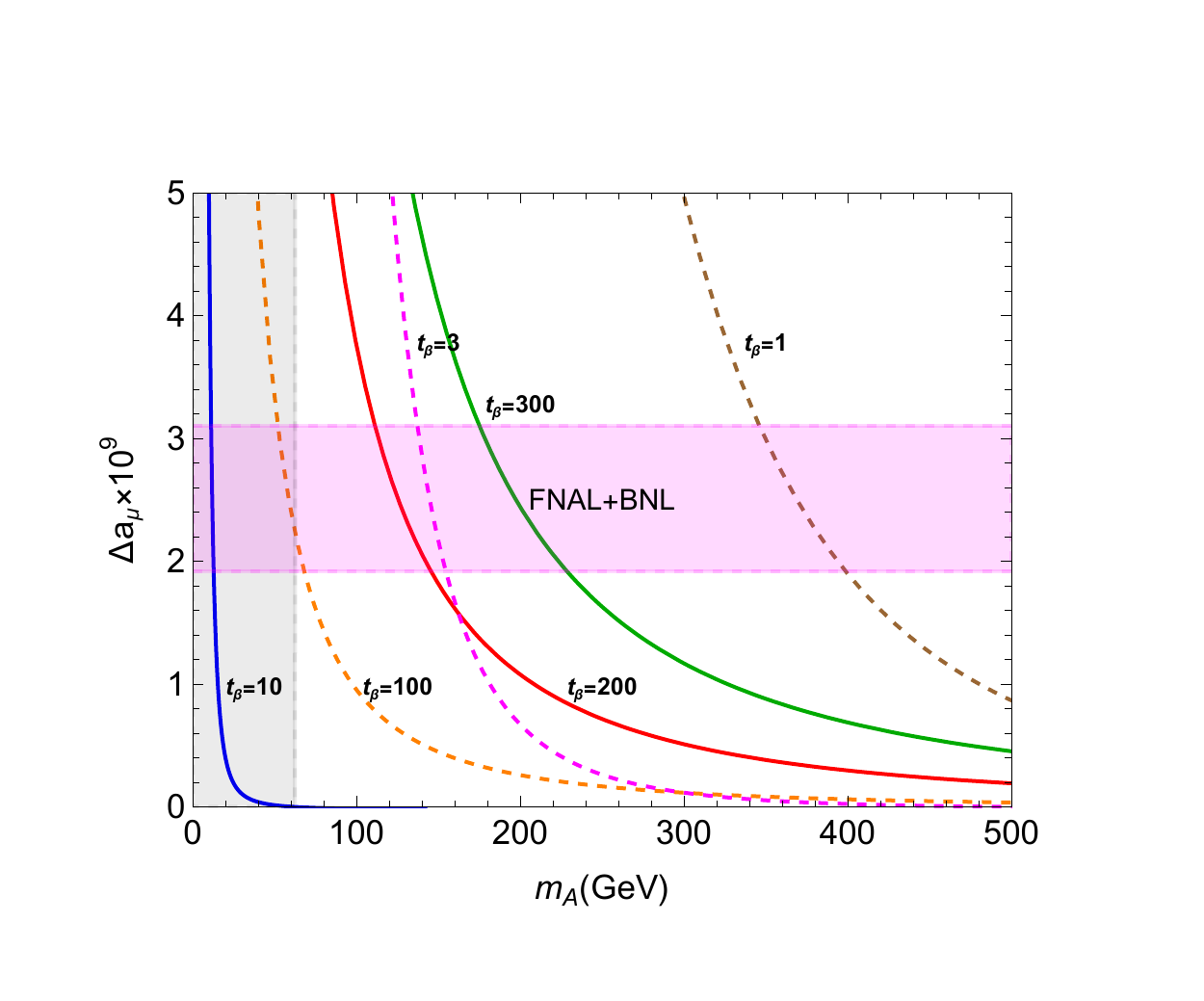}
    \vspace{-1cm}
    \caption{
    For $\kappa_{d}=\kappa_{u}=v^2/v_\phi^2$, $m_{h_1}=125$~GeV, $\lambda_{d}-\lambda_{u}=5$,  $\lambda_{\phi}=0.01$, $v_{\phi}=10^9$~GeV, and different values of $t_{\beta}$, we show that the observed $\Delta a_\mu$ can be obtained, plotting as a function of $m_{A}$. The gray area shows the excluded region based on the bound $m_{A}\gtrsim m_{h_1}/2$~\cite{CMS:2018qvj}.}
    \label{mu beta}
\end{figure}

Having obtained analytical formulas for $g_l - 2$, we investigate possible solutions to the anomalous magnetic moment of the muon and for the recently measured $ \Delta a_\mu$ in Eq.~\eqref{delamu}.
We show that the observed $\Delta a_\mu$ can be obtained for a range of $t_{\beta}$ and small values of $\kappa$, $\kappa_{d}$, and $\kappa_{u}$, which is also consistent with the expansion of the scalar field mass spectra in terms of $v/v_\phi$.
In Fig.~\ref{mu beta}, we fix $\kappa_{d}=\kappa_{u}=v^2/v_\phi^2$, $m_{h_1}=125$~GeV, $\lambda_{\phi}=0.01$, $\lambda_{d}-\lambda_{u}=5$, $v_{\phi}=10^9$~GeV and obtain $\Delta a_\mu$ for a range of $\kappa$ and some representative values of $t_{\beta}$ that simultaneously determine the value of $m_{h_3}$, $m_{h_2}$ and $m_{A}$, showing in terms of $m_{A}$.

As it can be seen in Fig.~\ref{mu beta}, we can find the observed anomaly for a range of $t_{\beta}$ values. Taking into account the bounds from colliders on $m_{A}\gtrsim m_{h_1}/2$~\cite{CMS:2018qvj}, in addition to the small $t_{\beta}$, we can find the observed $\Delta a_\mu$ for $ t_{\beta}\gtrsim 100$ \footnote{In some type of 2HDMs, the anomalous muon magnetic moment can be explained for large $t_{\beta}$ and a light pseudoscalar Higgs~\cite{Broggio:2014mna, Jueid:2021avn}.}.
It should be noted that for $1\lesssim t_{\beta}\lesssim 5$ the two-loop Barr-Zee contribution is dominated, whereas for other values of $t_{\beta}$ the one- and two-loop contributions, which are at the same order in $\tilde{y}_l$, are almost equivalent. Additionally, in the small values of $t_{\beta}$, the top quark in the two-loop diagram contributes predominantly, while in the large values of $t_{\beta}$, the dominant contribution comes from the bottom quark. Furthermore, with $\lambda_{d}-\lambda_{u}\lesssim 0.2$, the result of interest is obtained only for large $t_{\beta}$.

\subsection[Electron $g_e - 2$ anomaly]{\boldmath Electron $g_e - 2$ anomaly}
As already mentioned, due to the discrepancy in the measurement of $\alpha$ of Berkeley and LKB experiments, two different values have been reported for $\Delta a_e$, with opposite signs. Although results of the two experiments are rather inconsistent, we also study the averaged value case, combing the two results \cite{Botella:2022rte}
\begin{equation}\label{avg}
	\Delta a_e^{\textrm{Avg}}=(-2\pm 2.2)\times 10^{-13}.
\end{equation}
We first try to explore possible explanations for $\Delta a_e^{\textrm{LBK}}$ and $\Delta a_e^{\textrm{Avg}}$. We then study the physics which leads to the explanation of $\Delta a_e^{\textrm{B}}$, considering the constraints imposed by charged lepton flavor violation decays.

\subsubsection[$\Delta a_e^{\textrm{LBK}}$]{\boldmath $\Delta a_e^{\textrm{LBK}}$} \label{pe}
\begin{figure}
	\centering
	\includegraphics[scale=1]{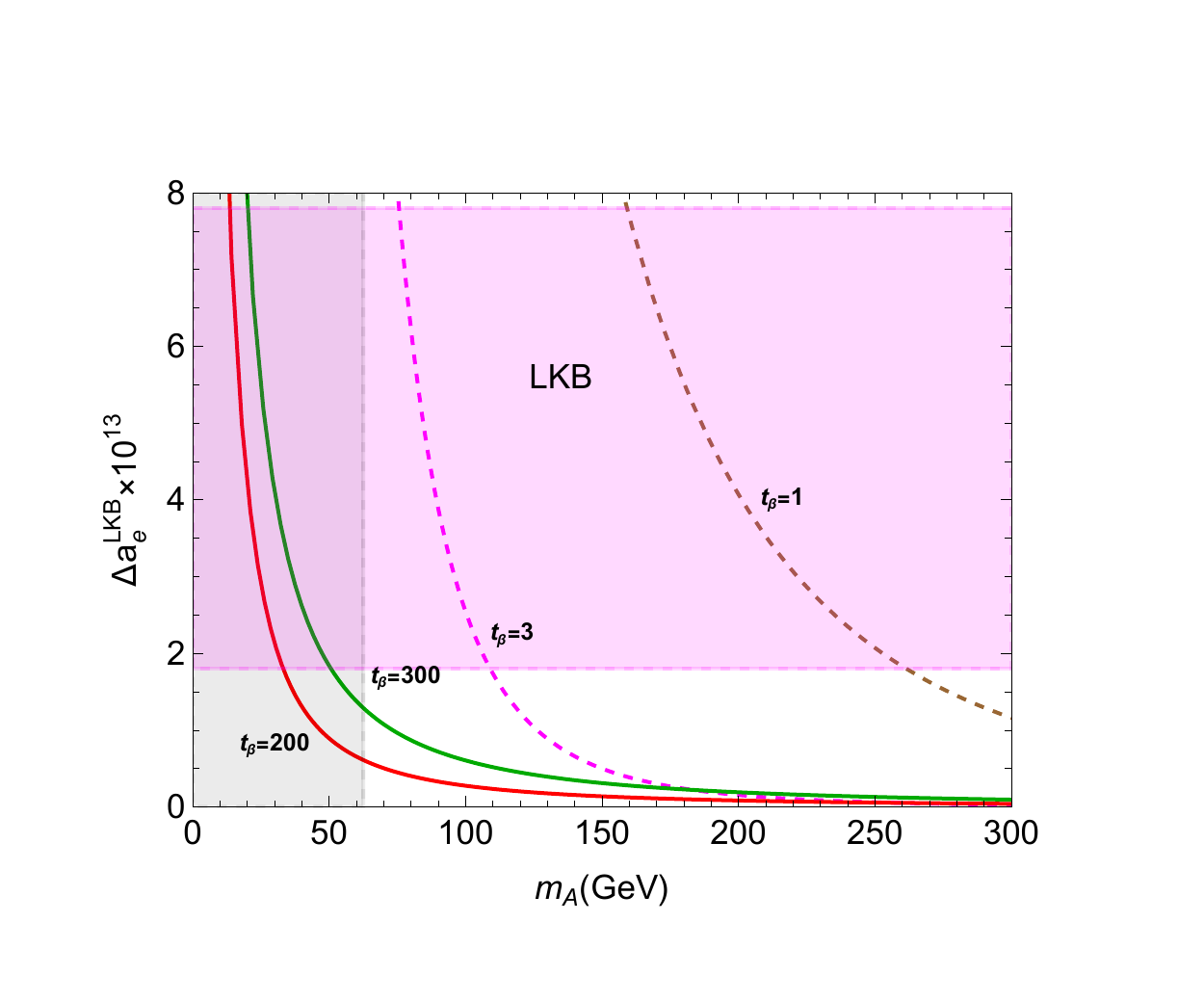}
        \vspace{-1cm}
	\caption{Similar to values chosen for parameters in the Fig.~\ref{mu beta}, for different values of $t_{\beta}$, we show that the observed $\Delta a_e^{\textrm{LBK}}$ can be obtained with respect to $m_{A}$.}
	\label{e beta}
\end{figure}

According to the LKB experiment and its result, Eq.\ (\ref{lkb}), and because of the muon and electron mass difference, $m_e/m_{\mu}\sim 5\times 10^{-3}$, it is expected to obtain positive $\Delta a_e$ with the similar benchmark chosen for the case $\Delta a_\mu$. Therefore, as shown in Fig.~\ref{e beta}, we can find that $\Delta a_e^{\textrm{LBK}} $ can be obtained as a function of $m_A$ and $t_{\beta}$. It is shown that $\Delta a_e^{\textrm{LBK}} $ can be explained for $1\lesssim t_{\beta}\lesssim 3$ where the two-loop Barr-Zee contribution is dominated, while for large $ t_{\beta}$, this is fulfilled for $ t_{\beta}\gtrsim 500$. In these regions of the parameter space, based on our previous discussion,  both the $W$-boson mass and $\Delta a_\mu$ anomalies can be explained. However, as can be seen from Figs.~\ref{e beta}, \ref{mu beta}, the range of $m_A$ values explaining $\Delta a_e^{\textrm{LBK}} $ and $\Delta a_\mu$ is different and thus both cannot be simultaneously explained. 
 
\begin{figure}
	\centering
	\includegraphics[scale=1]{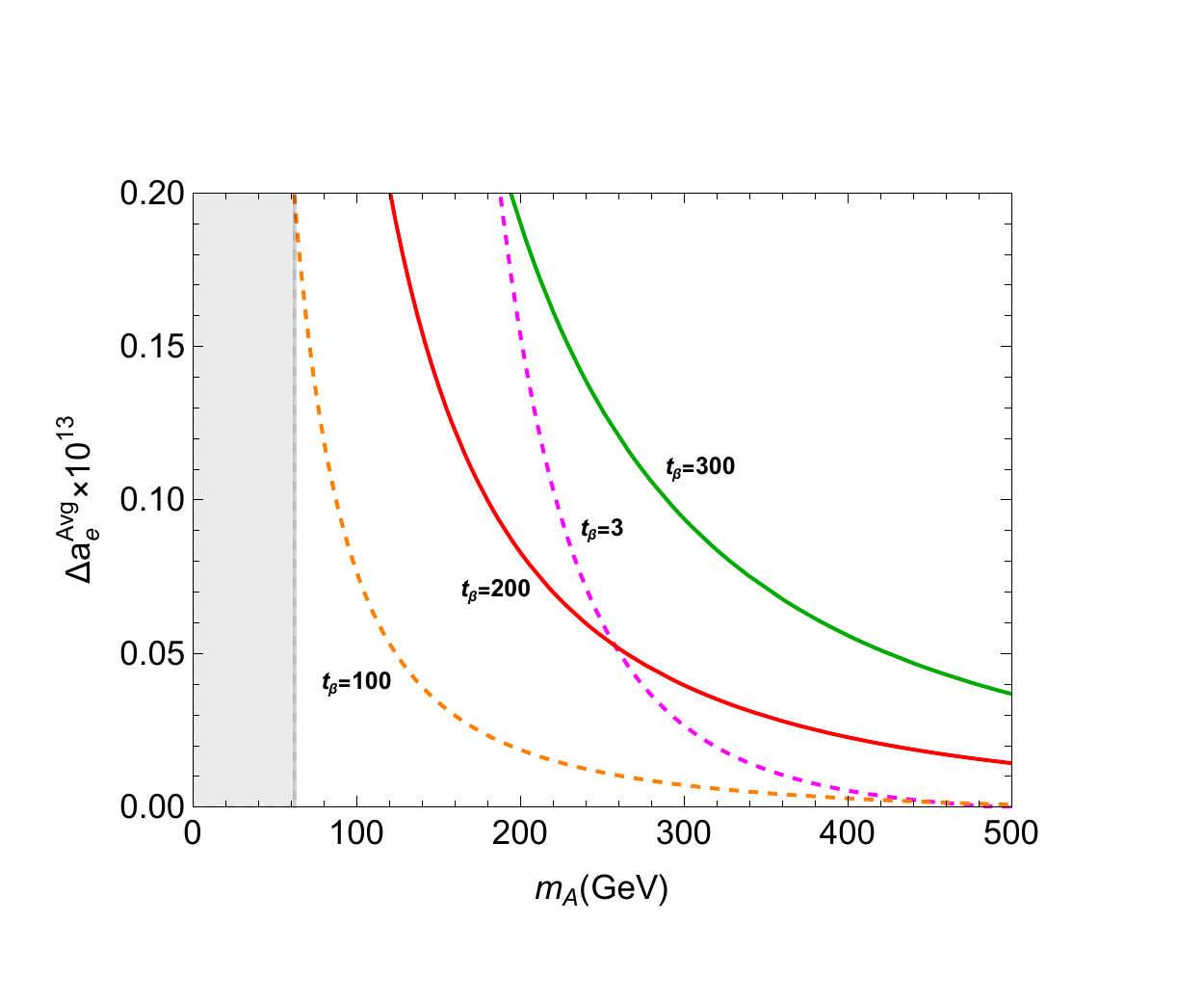}
	\vspace{-1cm}
	\caption{Similar to values chosen for parameters in the Fig.~\ref{mu beta}, for different values of $t_{\beta}$, we show that $\Delta a_e^{\textrm{Avg}}$ can be obtained, plotting as a function of $m_{A}$. The whole range of $\Delta a_e^{\textrm{Avg}}$ axis in this plot is allowed based on the average experimental bound. Because of the obtained positive results, negative values of Eq.\ (\ref{avg}) in the $\Delta a_e^{\textrm{Avg}}$ axis are not shown.
}
	\label{e avg}
\end{figure}

\begin{figure}
	\centering
	\includegraphics[scale=0.39]{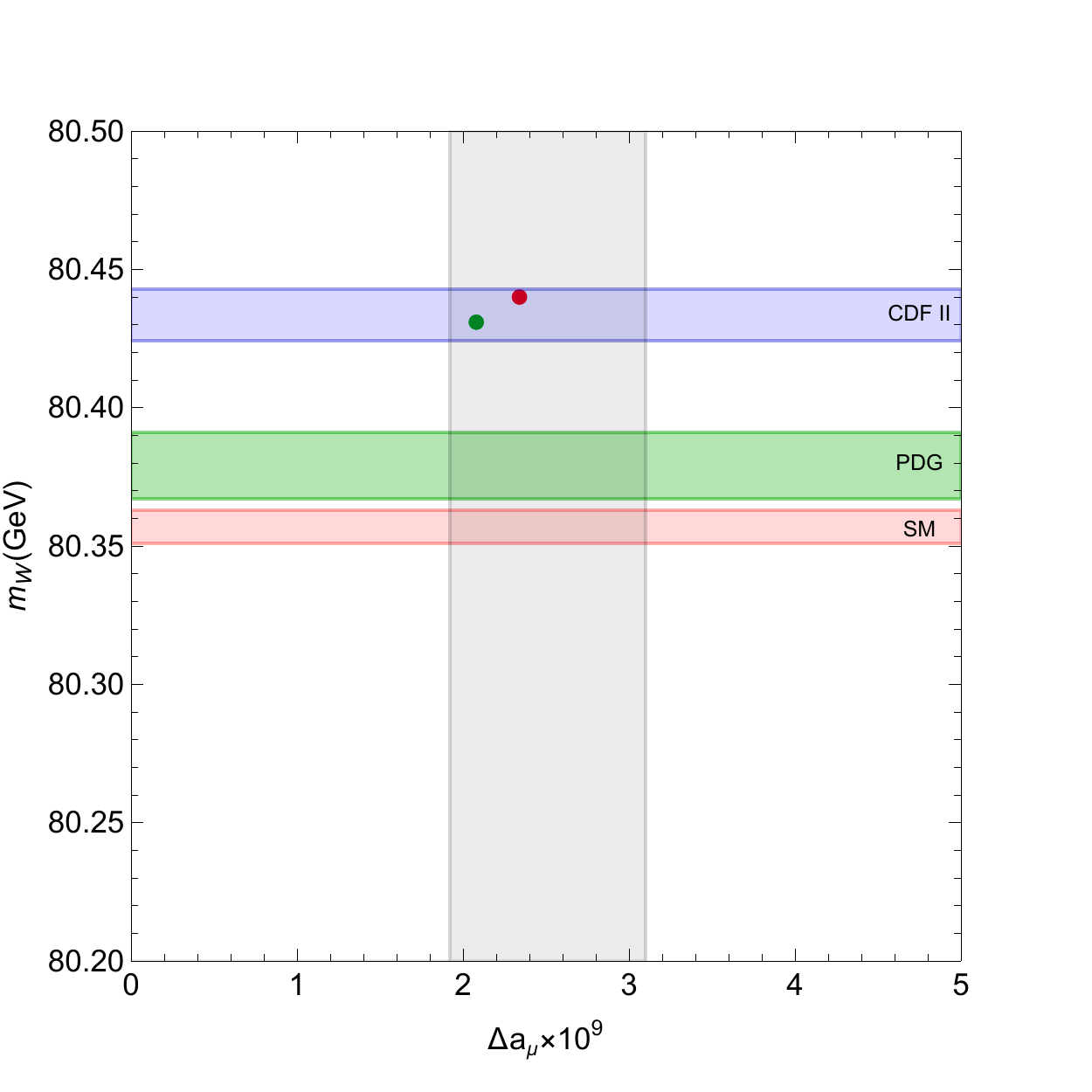}
	\includegraphics[scale=0.39]{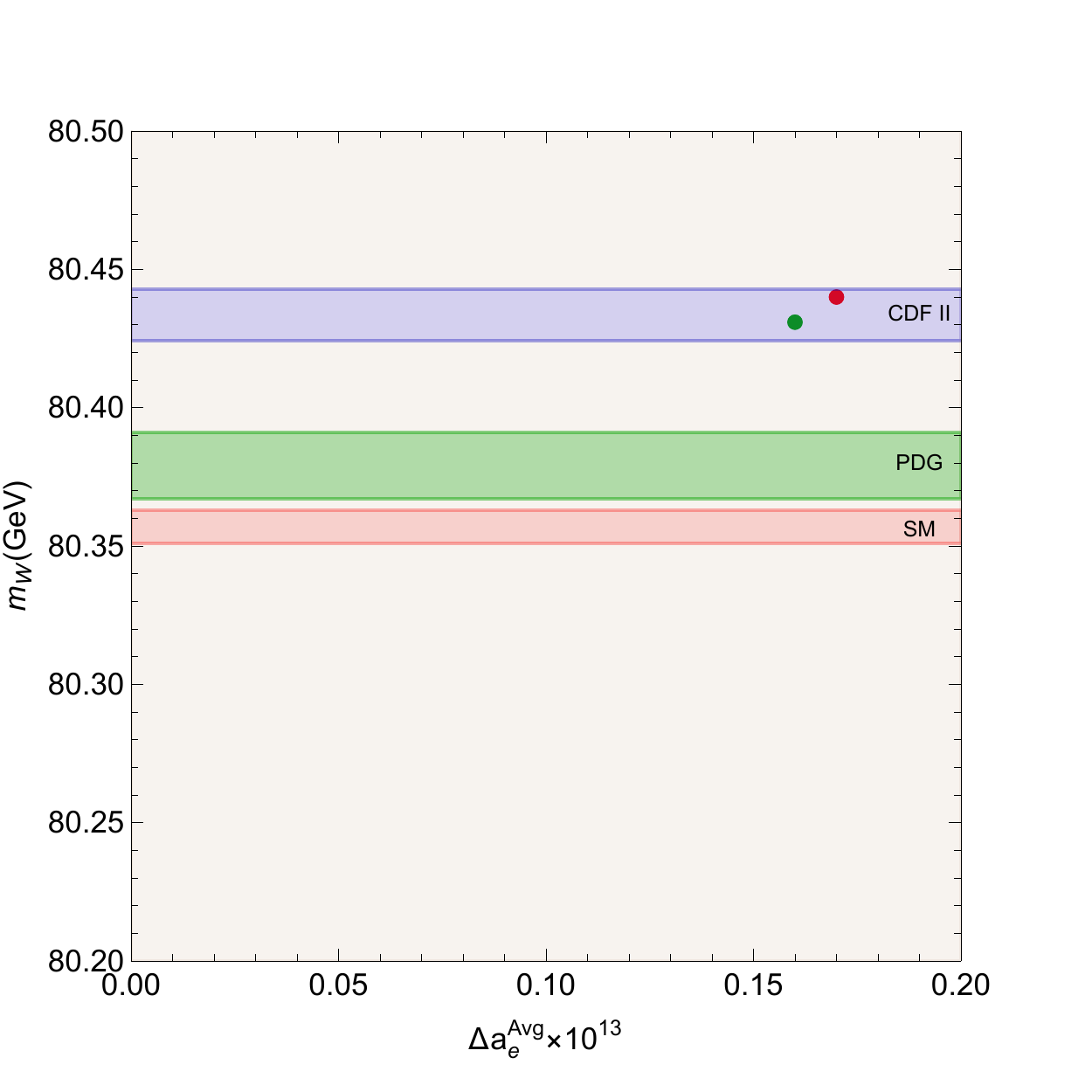}
	\includegraphics[scale=0.39]{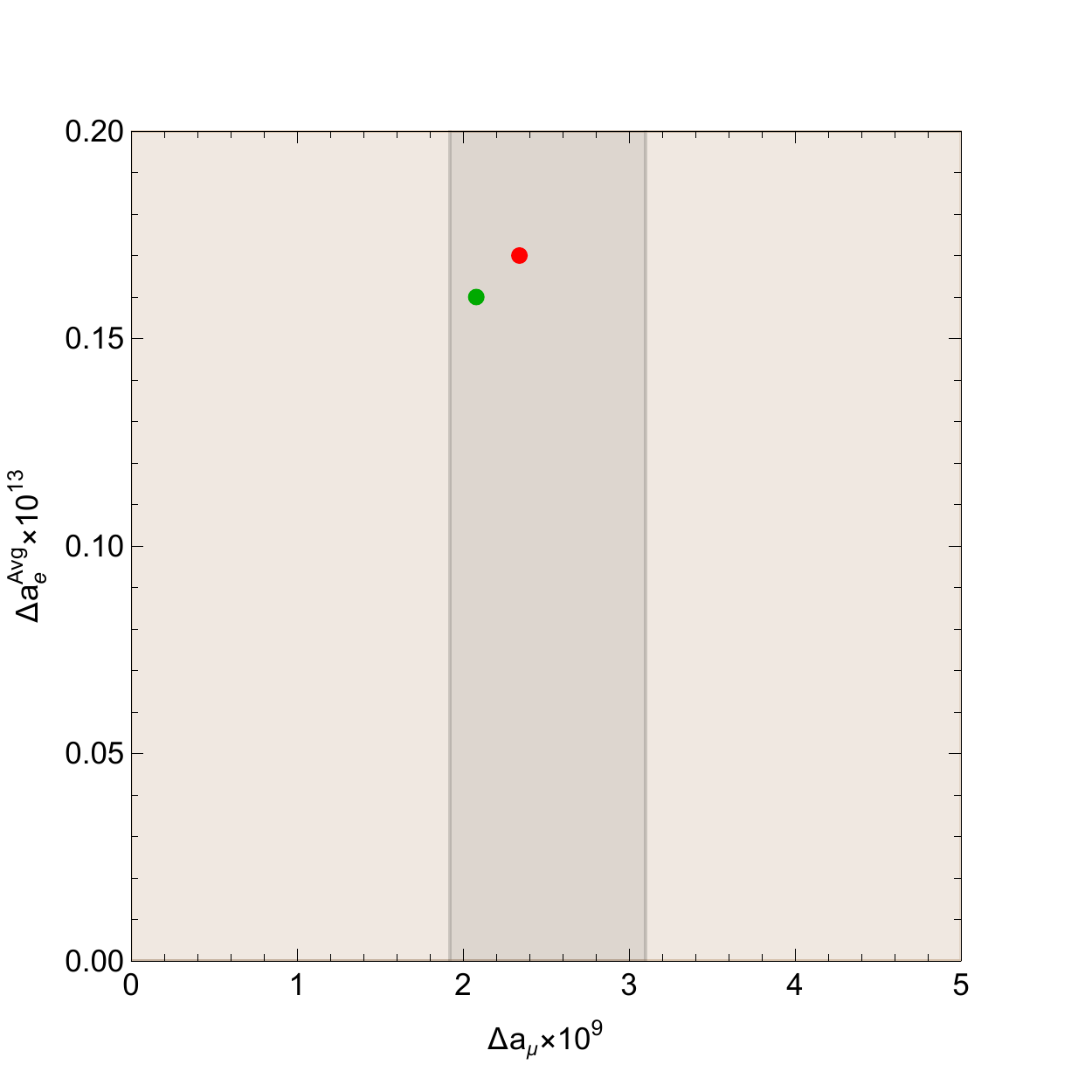}
        \vspace{0cm}
	\caption{Benchmark points, associated with $t_{\beta}=200$ (red) and $t_{\beta}=300$ (green), are represented, satisfying different experimental bounds for $M_{\rm W}$, $\Delta a_{\mu}$ and $\Delta a_e^{\textrm{Avg}}$ discussed in previous sections. Also, different experimental constraints are shown in three panels.} 
	\label{fig:bench}
\end{figure}

As shown in Fig.~\ref{e avg}, we also find that $\Delta a_e^{\textrm{Avg}}$ can be obtained within the parameter space of the model. Interestingly, given the averaged value measured for the electron magnetic moment, in addition to the electron magnetic moment, the $W$-boson mass, and muon magnetic moment anomalies can be simultaneously explained for $t_{\beta}\gtrsim 120$. For instance with the following benchmark, for $\kappa_{d} = \kappa_{u} = v^2 /v_\phi^2$, $m_{h_1}=125$~GeV, $\lambda_{d}-\lambda_{u}=5$,  $\lambda_{\phi}=0.01$, $v_{\phi}=10^9$~GeV, $ \lambda=0.048$, $\tilde{\kappa}=0.00035$ and $t_{\beta}=200$, one can obtain $m_W\simeq 80.44$~GeV, $\Delta a_\mu\simeq 2.34\times 10^{-9}$, and  $\Delta a_e^{\textrm{Avg}}\simeq 0.17\times 10^{-13}$. Also, as another example with the same parameters except for $\tan \beta =300$ and $\tilde{\kappa}= 0.0007$, we find $m_W\simeq 80.43$~GeV, $\Delta a_\mu\simeq 2.08\times 10^{-9}$, and  $\Delta a_e^{\textrm{Avg}}\simeq 0.16\times 10^{-13}$.
  These benchmark points are shown in Figure \ref{fig:bench} with red and green color, respectively.

\subsubsection[$\Delta a_e^{\textrm{B}}$]{\boldmath $\Delta a_e^{\textrm{B}}$} \label{ne}
As we discussed in the previous section, the parameter space of the model in which the $W$-boson mass and $\Delta a_\mu$ anomalies can be explained gives rise to a positive electron $g-2$. Therefore, to explain the case of negative electron magnetic moment, Eq.\ (\ref{ber}), the mentioned model should be extended.  Here we consider the model supplemented with heavy RH neutrinos, which are PQ charged and SM singlet, and study their contribution to the lepton $g_l - 2$.

The Lagrangian for right-handed neutrinos is given by 
\begin{equation} \label{lagn}
    \mathcal{L_{N}}= \overline{N}_{R_{\alpha}}i\partial_\mu N_{R_{\alpha}}-m_{N_{R_{\alpha}}}\overline{N}_{R_{\alpha}} N_{R_{\alpha}}^c-	y_{N_{\alpha}}^l\overline{\Psi}_{L_l} \widetilde{H}_d N_{R_{\alpha}}+\text{H.c.} \,,
\end{equation}
where $N_{R_{\alpha}} $ are RH neutrinos. Then, the Yukawa interaction of the charged Higgs and RH neutrinos can be expressed as\footnote{Here, we do not discuss the mechanism describing light neutrino masses and consider $y_{N_{\alpha}}^l$ as a variable with its diagonal form.}
\begin{equation} \label{yukn1}
    \sqrt{2}\, t_{\beta}\, y_{N_{\alpha}}^l\, H_{+}\, \overline{l}\, P_{R}\, N_{\alpha}+ \text{H.c.} \,.
\end{equation}
\begin{figure}
	\centering
	\includegraphics[scale=1.7]{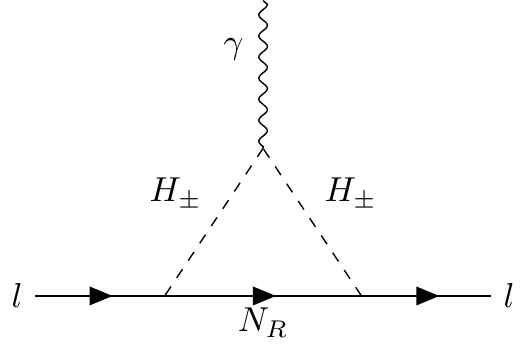}
	\caption{One-loop Feynman diagrams contributing to $(g_l-2)$ from charged Higgs fields and heavy neutrinos.}
	\label{feyn3}
\end{figure}
Therefore, according to the relevant Feynman diagram, Fig.~\ref{feyn3}, an additional one-loop amplitude should be calculated and (checked by using {\tt Package-X}~\cite{Patel:2015tea}) the result would be as follows
\begin{equation}\label{1lpn}
    \Delta a_{l}^{1L} = \sum_{\alpha=1}^{n_R} \frac{t_{\beta}^{2}\,\left(y_{N_{\alpha}}^l\right)^{2}}{8\pi^{2}}F_{H^{\pm}}(x_{H^{\pm}},z_{H^{\pm}})\,,
\end{equation}
where $x_{H^{\pm}}=m_{l}/m_{H^{\pm}}$, $z_{H^{\pm}}=m_{N_{R_{\alpha}}}/m_{H^{\pm}}$, and $n_R$ is the number of RH neutrinos. In our calculations, we assume $n_R=3$ and the same properties for all generations. 

\begin{figure}
    \centering
    \includegraphics[scale=.8]{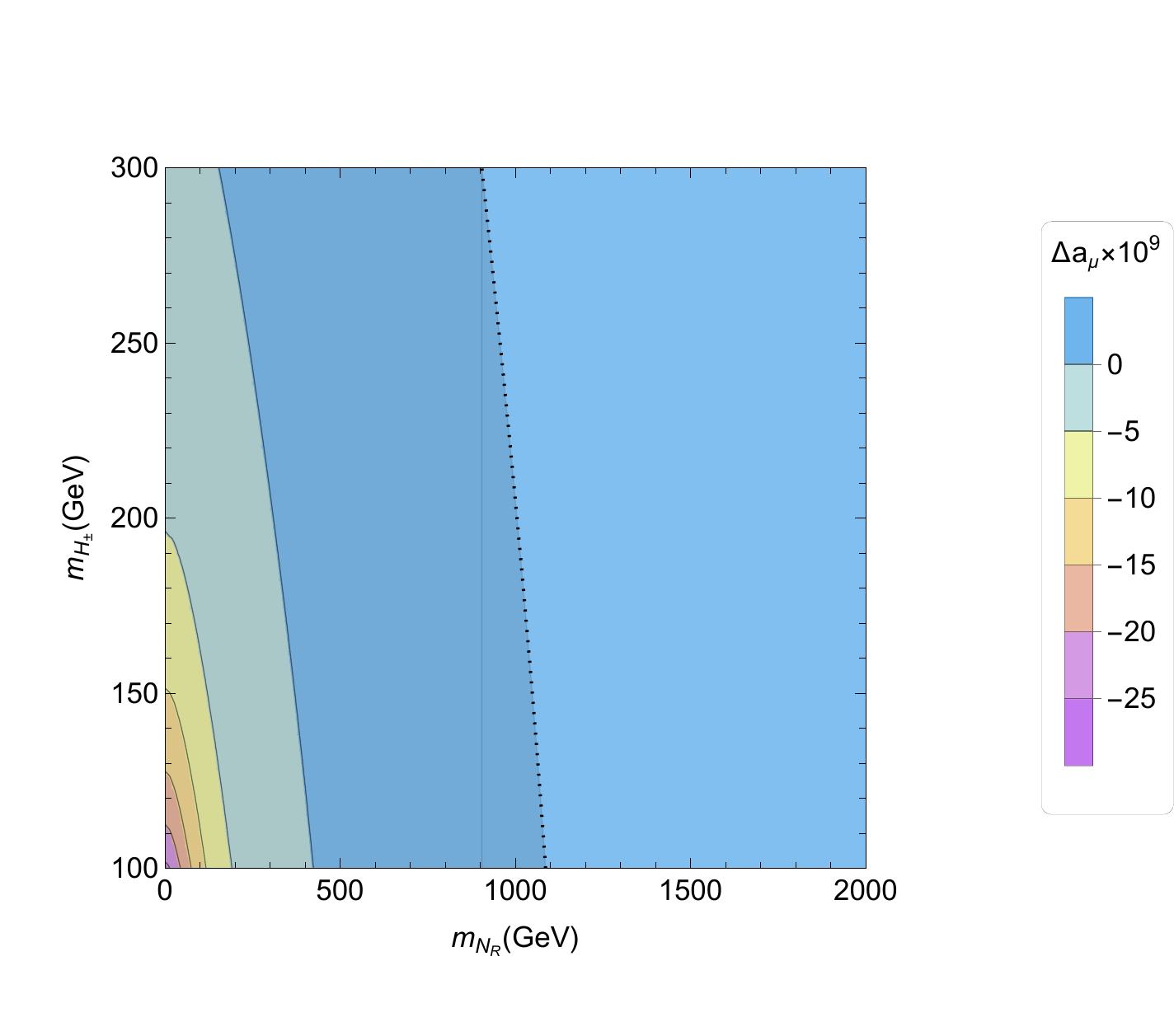}
    \vspace{-1cm}
    \caption{
    For $\kappa_{d}=\kappa_{u}=v^2/v_\phi^2$, $m_{h_1}=125$~GeV, $\lambda_{\phi}=0.01$, $\lambda_{d}-\lambda_{u}=5$, $v_{\phi}=10^9$~GeV, $t_{\beta}= 200$, $ m_{A}=130$~GeV and $y_N^{\mu}= 0.01$, we find $\Delta a_\mu$ as a function of $m_{N_{R}}$ and $m_{H^{\pm}}$. The colored regions show different values of the muon magnetic moment which could be even negative for small values of $m_{N_{R}}$ with $y_N^{\mu}= 0.01$. For $y_N^{\mu}< 0.01$ the negative contribution, Eq.~\eqref{1lpn}, to $\Delta a_\mu$ would be negligible. The region on the left of dashed grey line is excluded by the PDG $(g-2)_{\mu}$ bound.}
    \label{mu neut}
\end{figure}

Since this additional contribution, Eq.~\eqref{1lpn}, is negative, its effect on the muon magnetic moment should also be noted. For very heavy RH neutrinos, the contribution of this term is negligible. However, as shown in Fig.~\ref{mu neut}, for $y_N^{\mu}\gtrsim 0.01$ and sufficiently light RH neutrinos, $\Delta a_\mu$ can be negative, and hence we consider it $y_N^{\mu}< 0.01$ in the calculations.

\begin{figure}
    \centering
    \includegraphics[scale=.8]{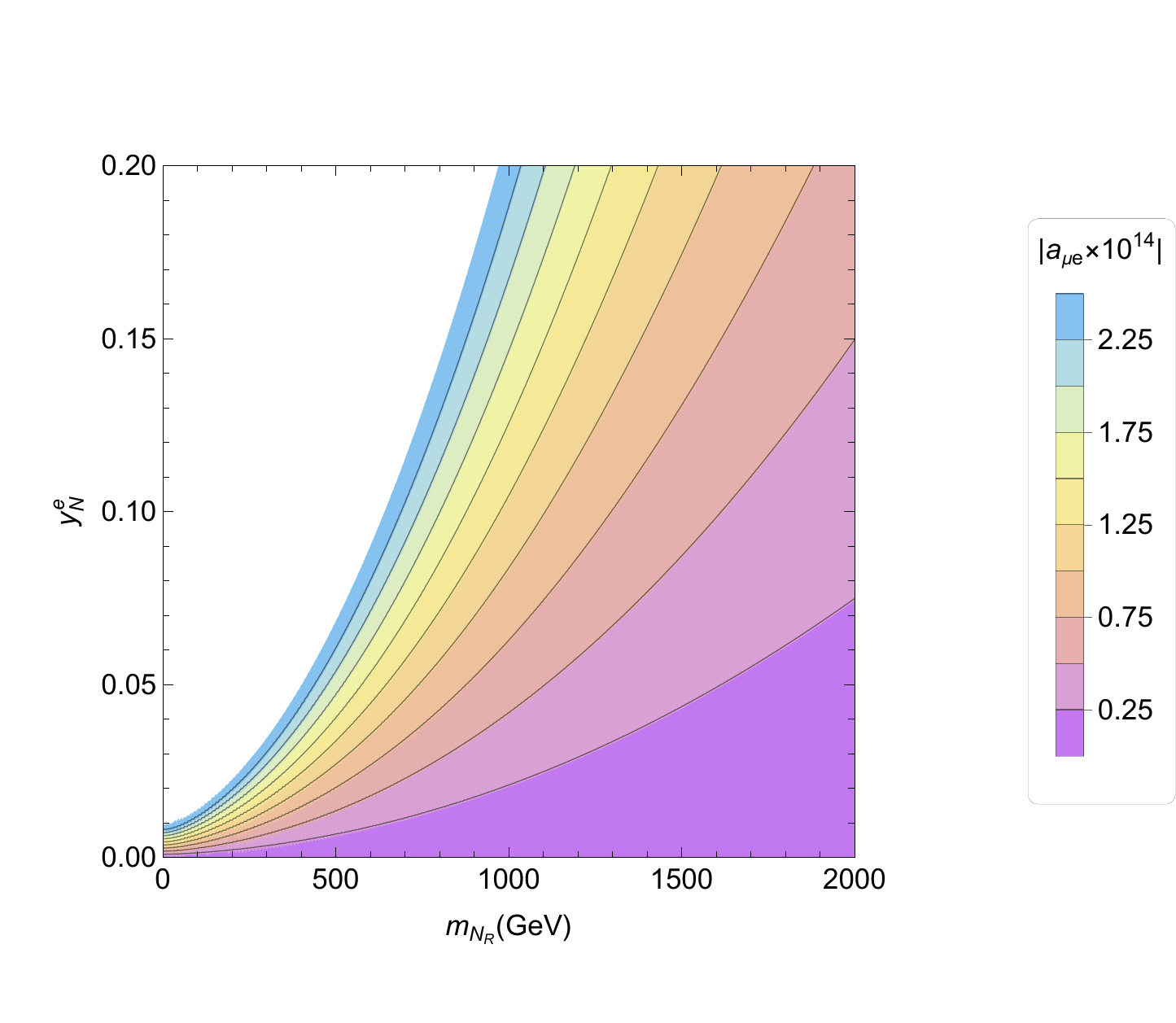}
    \vspace{-1cm}
    \caption{
    For $\kappa_{d}=\kappa_{u}=v^2/v_\phi^2$, $m_{h_1}=125$~GeV, $\lambda_{\phi}=0.01$, $\lambda_{d}-\lambda_{u}=5$, $v_{\phi}=10^9$~GeV, $t_{\beta}= 200$, $ m_{A}=130$~GeV, $ m_{H^{\pm}}=160$~GeV, and $y_N^{\mu}= 10^{-7}$, we show that $|a_{\mu e}|<2.5\times 10^{-14}$ can be obtained, plotting as a function of $m_{N_R}$ and $y_N^{e}$. All colored regions are consistent with the bound. }
    \label{lfv}
\end{figure}

\begin{figure}
    \centering
    \includegraphics[scale=.8]{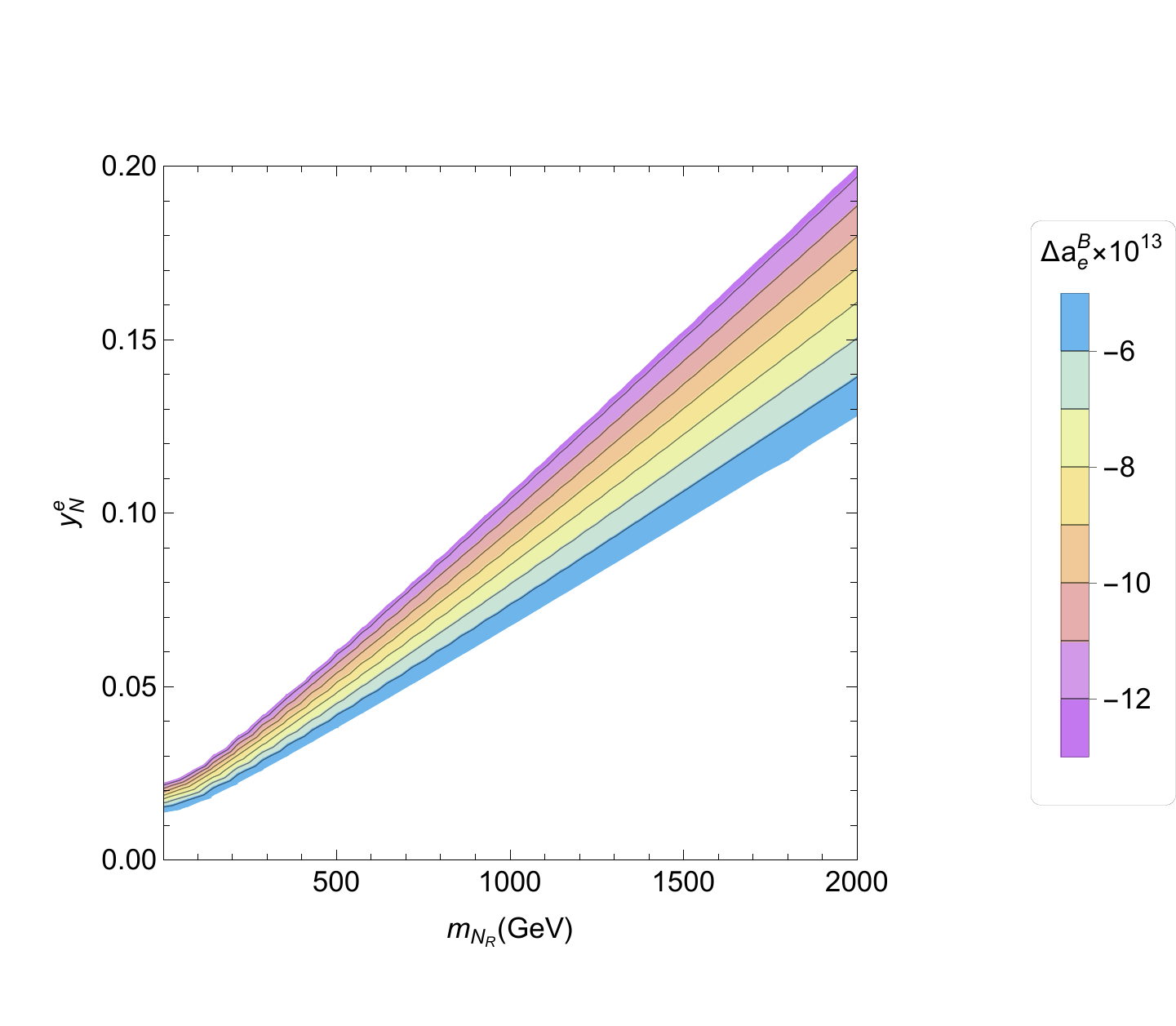}
    \vspace{-1cm}
    \caption{
    For $\kappa_{d}=\kappa_{u}=v^2/v_\phi^2$, $m_{h_1}=125$~GeV, $\lambda_{\phi}=0.01$, $\lambda_{d}-\lambda_{u}=5$, $v_{\phi}=10^9$~GeV, $t_{\beta}= 200$, $ m_{A}=130$~GeV, and $ m_{H^{\pm}}=160$~GeV, $\Delta a_e^{\textrm{B}}$ as a function of $m_{N_{R}}$ and  $y_{N}^{e}$ is shown. The colored regions show different negative values of the electron magnetic moment. }
    \label{B elec g-2}
\end{figure}

Before proceeding to calculate the electron magnetic moment for this case, we first discuss another constraint. Charged lepton flavor violations are highly constrained and the bound on the branching ratio of the decay $\mu\rightarrow e\gamma$ is more severe~\cite{MEG:2016leq}
\begin{equation}\label{fvbound}
    \mathrm{Br}(\mu\rightarrow e\gamma)<4.2\times 10^{-13}\,.
\end{equation}
Thus, we can constrain the model as follows
\begin{equation}
    \mathrm{Br}(\mu\rightarrow e\gamma)\sim \frac{96\pi^3\alpha\, v^4}{m_\mu^4}a_{\mu e}^2\,,
\end{equation}
where
\begin{equation}
    a_{\mu e}=\sum_{\alpha=1}^{n_R} \frac{t_{\beta}^{2}\,y_{N_{\alpha}}^{\mu}y_{N_{\alpha}}^e}{8\pi^{2}}\mathcal{F}_{H^{\pm}}\,,
\end{equation} 
and the function obtained $\mathcal{F}_{H^{\pm}} $ is given in Appendix \ref{app-loop}.\footnote{Note that for this constraint, we consider the case where the coupling of RH neutrinos and muon (muon neutrino), $y_{N_{\alpha}}^{\mu}$, is non-zero.}
According to Eq.~\eqref{fvbound}, we should have $|a_{\mu e}|<2.5\times 10^{-14}$. To pass the flavor violation decay bound, based on the previous estimation on $y_N^{\mu}< 0.01$, with calculating $a_{\mu e}$, the hierarchy $y_N^{\mu}\ll y_N^{e}$ can be imposed, taking $ y_N^{e}< 4\pi$. For example, in Fig.~\ref{lfv}, we show that $|a_{\mu e}|<2.5\times 10^{-14}$ can be obtained for $ y_N^{e}\leq 0.2$ and $y_N^{\mu}= 10^{-7}$.
Considering such bounds on the parameters, we obtain the electron magnetic moment and if $\Delta a_e^{\textrm{B}}$ is the case, we find that taking $ y_N^{e}<4\pi$, the measured quantity can be obtained for $  m_{N_R}\lesssim 200$~TeV. In Fig.~\ref{B elec g-2}, we show the result for $ y_N^{e}\leq 0.2$. 

\section{Conclusion} \label{sec:conc}

Recent measurements of the muon magnetic moment, the $W$-boson mass, and their deviation from the SM prediction can imply new physics beyond the SM. In this paper, in light of these observations we have studied a QCD axion model, enjoying the PQ symmetry, as an attractive framework 
that can address other SM shortcomings including the strong CP and the DM problem. 

We considered the DFSZ axion model and calculated loop-level diagrams that contribute to the muon and the electron magnetic moment, and $W$-boson mass. Considering the lower bound on the mass of charged Higgs, we have shown that the CDF-II $W$ boson mass excess can be addressed within the parameter space of the model, except for the case $m_{H^{\pm}}= m_A$, $ m_{h_2} $. We have also shown that the positive muon magnetic moment $\Delta a_\mu$ can be obtained for a range of $ t_{\beta}$ values with a mass range of extra Higgs bosons at the EW scale, respecting the direct collider bound on the pseudoscalar field and also theoretical constraints. 

Regarding the electron $g-2$, there are Berkeley and LKB experiments whose measurements are set in opposite directions. We showed that $\Delta a_e^{\textrm{LBK}} $ can be explained within a different region of the parameter space allowed for $\Delta a_\mu$. However, we found that considering the averaged value of the two experiments, the three anomalies can be simultaneously explained. The electron magnetic moment measured by another experiment at Berkeley, $\Delta a_e^{\textrm{B}}$, is of a different sign and the physics that explains $\Delta a_\mu$ and the $W$-boson mass excess leads to a positive electron $g-2$. To address $\Delta a_\mu$, $\Delta a_e^{\textrm{B}}$, and the $W$-boson mass anomalies simultaneously, we extended the model with heavy RH neutrinos. Taking into account the flavor violation decay bounds, we found that heavy neutrino-charged Higgs interactions contributing to one-loop diagrams that can be responsible for the negative $\Delta a_e$ with heavy neutrinos masses up to 200 TeV.

\acknowledgments
The authors would like to thank Niclo{\'a}s Bernal for  insightful discussions and comments. Also, they are grateful to Heinrich P{\"a}s for reading the early version of our manuscript.  
FH is supported by the research grant ``New Theoretical Tools for Axion Cosmology'' under the Supporting TAlent in ReSearch@University of Padova (STARS@UNIPD), Istituto Nazionale di Fisica Nucleare (INFN) through the Theoretical Astroparticle Physics (TAsP) project. FH also thanks the organizers of the ``Probing New Physics with Gravitational Waves'' Workshop 2022 at the Mainz Institute for Theoretical Physics (MITP) of the Cluster of Excellence PRISMA (Project ID 39083149) in Mainz, for hospitality and partial financial support during the completion of this work.

\appendix

\section{Minimization and vacuum stability conditions}\label{min}
We can use  the minimization of the potential, Eq.~\eqref{pot}, and remove three parameters $\mu_{d} $, $\mu_{u} $ and $V_{\phi} $, expressing in terms of free parameters as follows
\begin{align}
    2\lambda_{d} v^{2} c_{\beta}^{2}&-\mu_{d}^{2}+\frac{v_{\phi}^{2}}{2}\left(\kappa_{d}-\frac{\kappa \tan \beta}{v_{\phi}}\right)=0\,, \\
	2\lambda_{u} v^{2} s_{\beta}^{2}&-\mu_{u}^{2}+\frac{v_{\phi}^{2}}{2}\left(\kappa_{u}-\frac{\kappa}{v_{\phi}\tan \beta}\right)=0\,, \\
	\lambda_{\phi}\left(v_{\phi}^{2}-V_{\phi}^{2}\right)&+2 v^{2}\left(\kappa_{d} c_{\beta}^{2}+\kappa_{u} s_{\beta}^{2}-\frac{\kappa s_{2 \beta}}{v_{\phi}}\right)=0 \,.
\end{align}
Assuming that $\kappa_{d}$, $\kappa_{u}$ and $\kappa$ are sufficiently small and do not affect the vacuum stability conditions obtained from other quartic terms of the potential, we have
\begin{equation}
	\begin{aligned}
		&\lambda_{\phi}>0\,, \quad \lambda_{d}>0\,, \quad \lambda_{u}>0 \,,\\
		&\lambda+2\sqrt{\lambda_{d}\lambda_{u}}>0 \,.
	\end{aligned}
\end{equation}

\section{
Rotations for the Higgs fields} \label{rot}
With the following rotated Higgs doublets, one can find the direction by which unphysical components are removed
\begin{equation}
	H_d' = c_\beta H_{d}+s_{\beta} H_{u} = \frac{1}{\sqrt{2}} \left(
        \begin{array}{c}
		c_{\beta}\, \alpha^{+} + s_{\beta}\, \beta^{+} \\
		c_{\beta} \operatorname{Re} \left[\alpha_{0}\right] + s_{\beta} \operatorname{Re}\left[\beta_{0}\right] + i \left(c_{\beta} \operatorname{Im}\left[\alpha_{0}\right] + s_{\beta} \operatorname{Im}\left[\beta_{0}\right]\right)
	\end{array}
        \right),
\end{equation}
where $c_{\beta} \equiv \cos \beta$ and $s_{\beta} \equiv \sin \beta$.
The upper components are combinations absorbed by $W^\pm$ and the real part is the component with non-zero vev, thus $H_d'$ has the structure of SM doublet so that
\begin{equation}
	\langle c_{\beta} \operatorname{Re}\left[\alpha_{0}\right]+s_{\beta} \operatorname{Re}\left[\beta_{0}\right]\rangle =c_{\beta}v_d+s_{\beta}v_u=v
\end{equation}
and $\langle H+v\rangle=v$. Also, the other rotated doublet would be
\begin{equation}
	H_u' = -s_{\beta} H_{d}+c_{\beta} H_{u}=\frac{1}{\sqrt{2}}\left(\begin{array}{c}
		-s_{\beta} \alpha^{+}+c_{\beta} \beta^{+} \\
		-s_{\beta} \operatorname{Re}\left[\alpha_{0}\right]+c_{\beta} \operatorname{Re}\left[\beta_{0}\right]+i\left(-s_{\beta} \operatorname{Im}\left[\alpha_{0}\right]+c_{\beta} \operatorname{Im}\left[\beta_{0}\right]\right)
	\end{array}\right)\,,
\end{equation}
whose charged components denote the charged Higgs $H^{\pm}$ and the real neutral component is a scalar Higgs $S$ such that $\langle S\rangle=\langle -s_{\beta} \operatorname{Re}\left[\alpha_{0}\right]+c_{\beta} \operatorname{Re}\left[\beta_{0}\right]\rangle =0$. Therefore, the gauge-redefined doublets can be expressed as 
\begin{align} \label{transh}
		H_{d} &=c_{\beta} H_{d}^{\prime}-s_{\beta} H_{u}^{\prime}=\frac{1}{\sqrt{2}}\left(\begin{array}{c}
			-\sqrt{2} s_{\beta} H^{+} \\
			\left(v+H\right) c_{\beta}-s_{\beta}\left(S+i \tilde{A}\right)
		\end{array}\right) \,,\\
		H_{u}&=s_{\beta} H_{d}^{\prime}+c_{\beta} H_{u}^{\prime}=\frac{1}{\sqrt{2}}\left(\begin{array}{c}
			\sqrt{2} c_{\beta} H^{+} \\
			\left(v+H\right) s_{\beta}+c_{\beta}\left(S+i \tilde{A}\right)
		\end{array}\right).
\end{align}
Although $\tilde{A}$ remains after gauge redefinition, the field is not yet physical. Moreover, from $\phi=v_{\phi}+\varrho+i\tilde{G}$, $\tilde{G}$ is another non-physical d.o.f. whose combination with $\tilde{A}$ identifies the physical one in the pseudoscalar sector~\cite{Espriu:2015mfa}. In addition to the massless eigenstate, the axion resulting from the PQ current is
\begin{equation}
    a=\frac{ v_{\phi} \tilde{G}+v s_{2\beta} \tilde{A}}{\sqrt{ v_{\phi}^{2}+v^{2} s_{2\beta}^2}} \,.
\end{equation}
There is also a massive pseudoscalar eigenstate, evident from the cubic term in the potential, as
\begin{equation}\label{massa}
	A=\frac{ v_{\phi} \tilde{A}-v s_{2\beta} \tilde{G}}{\sqrt{ v_{\phi}^{2}+v^{2} s_{2\beta}^2}}\,.
\end{equation}

\section{
The mass matrix of neutral scalar fields}\label{diag}
The mass matrix of the neutral CP-even scalar fields is given by
\begin{equation}
M=\left(\begin{array}{ccc}
	32 v^2\left(\lambda_d c_\beta^4+\lambda_u s_\beta^4\right) & M_{12} & M_{13} \\
	M_{21} & \frac{8 \kappa v_\phi}{s_{2 \beta}}+8 v^2\left(\lambda_d+\lambda_u\right) s_{2 \beta}^2 & M_{23} \\
	M_{31} & M_{32} & 4\lambda_\phi v_\phi^2
\end{array}\right),
\end{equation}
where
\begin{equation}
    M_{12}=M_{21}=16 v^2\left(-\lambda_d c_\beta^2+\lambda_u s_\beta^2\right) s_{2 \beta}\,,
\end{equation}
\begin{equation}
    M_{13}=M_{31}=8 v v_\phi\left(\kappa_{d} c_\beta^2+\kappa_{u} s_\beta^2-\kappa s_{2 \beta}/v_\phi\right),
\end{equation}
\begin{equation}
    M_{23}=M_{32}=-4v v_\phi\left[(\kappa_{d}-\kappa_{u}) s_{2 \beta}+2 \kappa c_{2 \beta}/v_\phi\right].
\end{equation}

The rotation matrix can be written as~\cite{Espriu:2015mfa}
\begin{equation}
    R=\exp \left(\frac{v}{v_{\phi}} \mathcal{A}+\frac{v^{2}}{v_{\phi}^{2}} \mathcal{B}\right), \quad \mathcal{A}^{T}=-\mathcal{A}, \quad \mathcal{B}^{T}=-\mathcal{B} \,,
\end{equation}
so that
\begin{equation}
    R=\left(
    \begin{array}{ccc}
        1-\frac{v^{2}}{v_{\phi}^{2}} \frac{\mathcal{A}_{13}^{2}}{2} & \frac{v^{2}}{v_{\phi}^{2}} \frac{\mathcal{A}_{13} \mathcal{A}_{23}+2 \mathcal{B}_{12}}{2} & \frac{v}{v_{\phi}} \mathcal{A}_{13} \\
    -\frac{v^{2}}{v_{\phi}^{2}} \frac{\mathcal{A}_{13} \mathcal{A}_{23}+2 \mathcal{B}_{12}}{2} & 1-\frac{v^{2}}{v_{\phi}^{2}} \frac{\mathcal{A}_{23}^{2}}{2} & \frac{v}{v_{\phi}} \mathcal{A}_{23} \\
    -\frac{v}{v_{\phi}} \mathcal{A}_{13} & -\frac{v}{v_{\phi}} \mathcal{A}_{23} & 1-\frac{v^{2}}{v_{\phi}^{2}} \frac{\mathcal{A}_{13}^{2}+\mathcal{A}_{23}^{2}}{2}
    \end{array}
    \right) ,
\end{equation}
where
\begin{equation}
    \mathcal{A}_{13}=\frac{2}{\lambda_{\phi}}\left(\kappa_{d} c_{\beta}^{2}+\kappa_{u} s_{\beta}^{2}-\kappa s_{2 \beta}/v_{\phi}\right),\quad \mathcal{A}_{23}=\frac{(\kappa_{d}-\kappa_{u}) s_{2 \beta}+2 \kappa c_{2 \beta}/v_{\phi}}{\frac{2 \kappa}{v_{\phi}s_{2 \beta}}-\lambda_{\phi}}\,,
\end{equation}
\begin{align}
    \mathcal{B}_{12}&=-\frac{2v_{\phi}}{\kappa} s_{2 \beta}^{2}\left(\lambda_{d} c_{\beta}^{2}-\lambda_{u} s_{\beta}^{2}\right)\nonumber\\
    &\quad +\frac{v_{\phi}s_{2 \beta}}{\lambda_{\phi} \kappa} \frac{\kappa-v_{\phi}\lambda_{\phi} s_{2 \beta}}{2 \kappa-v_{\phi}\lambda_{\phi} s_{2 \beta}}\left(\kappa_{d} c_{\beta}^{2}+\kappa_{u} s_{\beta}^{2}-\frac{\kappa s_{2 \beta}}{v_{\phi}}\right)\left[(\kappa_{d}-\kappa_{u}) s_{2 \beta}+\frac{2 \kappa c_{2 \beta}}{v_{\phi}}\right],
\end{align}	
and up to second order in $v/v_{\phi}$, $ \mathcal{A}_{12}=\mathcal{B}_{13}=\mathcal{B}_{23}=0$. Thus, the scalar states are related to the mass eigenstates $h_i$, as
\begin{equation}
	\left(\begin{array}{c}
		H \\
		S \\
		\varrho
	\end{array}\right)=R\left(\begin{array}{l}
		h_{1} \\
		h_{2} \\
		h_{3}
	\end{array}\right).
\end{equation}
so that
\begin{equation}\label{massmat}
	H=\sum_{i=1}^{3} R_{H i} h_{i}, \quad S=\sum_{i=1}^{3} R_{S i} h_{i}, \quad \varrho=\sum_{i=1}^{3} R_{\rho i} h_{i} \,.
\end{equation}

\section{Loop functions} \label{app-loop}
Loop functions of one-loop calculations used in Section \ref{sec:gm2} are given as follows
\begin{align}
    &F(x) =\frac{\left(3 x^2-2\right) x^2+\left(3 x^2-1\right) \ln \left(x^2\right)+2 \left(x^2-1\right) \sqrt{1-4 x^2} \ln \left(\frac{1+\sqrt{1-4 x^2}}{2 x}\right)}{x^4}, \\
    &F_{H^{\pm}}(x,z) = \frac{x^2+2 z^2-2}{2 x^2}+\frac{\left(x^2-\left(z^2-1\right)^2\right) \log \left(z^2\right)}{2 x^4}\nonumber\\
    &-\frac{\left(x^4+x^2 \left(z^4+z^2-2\right)-\left(z^2-1\right)^3\right) \log \left(\frac{\left(-x^2+z^2+1\right)+\sqrt{-2 \left(x^2+1\right) z^2+\left(x^2-1\right)^2+z^4}}{2 z}\right)}{x^4 \sqrt{x^4-2 x^2 \left(z^2+1\right)+\left(z^2-1\right)^2}},
	\end{align}
and functions of the two-loop amplitude can be computed by
\begin{equation}
	\mathcal{F}\left(\omega\right)=\frac{\omega}{2}\int_0^1 d x\, \frac{2 x(1-x)-1}{x(1-x)-\omega} \log \frac{x(1-x)}{\omega}\,,
\end{equation}
\begin{equation}
	\tilde{\mathcal{F}}\left(\omega\right)=\frac{\omega}{2}\int_0^1 d x\, \frac{1}{x(1-x)-\omega} \log \frac{x(1-x)}{\omega}\,.
\end{equation}
Moreover, the function associated with the calculation of the lepton flavor violation process is obtained as
\begin{align}	
    &\mathcal{F}_{H^{\pm}}=\frac{\left(m_\mu m_e^2-m_\mu m_{H^{\pm}}^2+m_\mu m_N^2-2 m_e m_{H^{\pm}}^2+2 m_e m_N^2\right)  \Lambda \left(m_e^2,m_{H^{\pm}},m_N\right)}{4 m_e \left(m_\mu^2-m_e^2\right)} \nonumber\\
    &-\frac{\left(m_\mu^2 m_e-2 m_\mu m_{H^{\pm}}^2+2 m_\mu m_N^2-m_e m_{H^{\pm}}^2+m_e m_N^2\right) \Lambda \left(m_\mu^2,m_{H^{\pm}},m_N\right)}{4 m_\mu \left(m_\mu^2-m_e^2\right)} \nonumber\\
    &+\frac{1}{2} m_{H^{\pm}}^2 C_{0} \left(0,m_\mu^2,m_e^2,m_{H^{\pm}},m_{H^{\pm}},m_N\right)+\frac{m_\mu m_e-m_{H^{\pm}}^2+m_N^2}{4 m_\mu m_e} \nonumber\\
    &-\frac{\log \left(\frac{m_{H^{\pm}}^2}{m_N^2}\right) \left(2 m_\mu^2 m_e^2 m_{H^{\pm}}^2-m_\mu^2 m_{H^{\pm}}^4+2 m_\mu^2 m_{H^{\pm}}^2 m_N^2-m_\mu^2 m_N^4-2 m_\mu m_e m_{H^{\pm}}^4\right)}{8 m_\mu^3 m_e^3} \nonumber\\
    &-\frac{\log \left(\frac{m_{H^{\pm}}^2}{m_N^2}\right) \left(4 m_\mu m_e m_{H^{\pm}}^2 m_N^2-2 m_\mu m_e m_N^4-m_e^2 m_{H^{\pm}}^4+2 m_e^2 m_{H^{\pm}}^2 m_N^2-m_e^2 m_N^4\right)}{8 m_\mu^3 m_e^3},
\end{align}
where $\Lambda$ ({\tt DiscB}) and $C_{0}$ ({\tt ScalarC0}) are Veltmann-Passarino functions that are also defined in the {\tt Package-X} program.

\bibliographystyle{JHEP}
\bibliography{biblio}

\providecommand{\href}[2]{#2}\begingroup\raggedright\begin{thebibliography}{10}

\bibitem{Peccei:1977hh}
R.D.~Peccei and H.R.~Quinn, \emph{{CP Conservation in the Presence of
  Instantons}}, \href{https://doi.org/10.1103/PhysRevLett.38.1440}{\emph{Phys.
  Rev. Lett.} {\bfseries 38} (1977) 1440}.

\bibitem{Peccei:1977np}
R.D.~Peccei and H.R.~Quinn, \emph{{Some Aspects of Instantons}},
  \href{https://doi.org/10.1007/BF02730110}{\emph{Nuovo Cim. A} {\bfseries 41}
  (1977) 309}.

\bibitem{Peccei:2006as}
R.D.~Peccei, \emph{{The Strong CP problem and axions}},
  \href{https://doi.org/10.1007/978-3-540-73518-2_1}{\emph{Lect. Notes Phys.}
  {\bfseries 741} (2008) 3}
  [\href{https://arxiv.org/abs/hep-ph/0607268}{{\ttfamily hep-ph/0607268}}].

\bibitem{Weinberg:1977ma}
S.~Weinberg, \emph{{A New Light Boson?}},
  \href{https://doi.org/10.1103/PhysRevLett.40.223}{\emph{Phys. Rev. Lett.}
  {\bfseries 40} (1978) 223}.

\bibitem{Wilczek:1977pj}
F.~Wilczek, \emph{{Problem of Strong $P$ and $T$ Invariance in the Presence of
  Instantons}}, \href{https://doi.org/10.1103/PhysRevLett.40.279}{\emph{Phys.
  Rev. Lett.} {\bfseries 40} (1978) 279}.

\bibitem{Preskill:1982cy}
J.~Preskill, M.B.~Wise and F.~Wilczek, \emph{{Cosmology of the Invisible
  Axion}}, \href{https://doi.org/10.1016/0370-2693(83)90637-8}{\emph{Phys.
  Lett. B} {\bfseries 120} (1983) 127}.

\bibitem{Abbott:1982af}
L.F.~Abbott and P.~Sikivie, \emph{{A Cosmological Bound on the Invisible
  Axion}}, \href{https://doi.org/10.1016/0370-2693(83)90638-X}{\emph{Phys.
  Lett. B} {\bfseries 120} (1983) 133}.

\bibitem{Dine:1982ah}
M.~Dine and W.~Fischler, \emph{{The Not So Harmless Axion}},
  \href{https://doi.org/10.1016/0370-2693(83)90639-1}{\emph{Phys. Lett. B}
  {\bfseries 120} (1983) 137}.

\bibitem{Davis:1985pt}
R.L.~Davis, \emph{{Goldstone Bosons in String Models of Galaxy Formation}},
  \href{https://doi.org/10.1103/PhysRevD.32.3172}{\emph{Phys. Rev. D}
  {\bfseries 32} (1985) 3172}.

\bibitem{Marsh:2015xka}
D.J.E.~Marsh, \emph{{Axion Cosmology}},
  \href{https://doi.org/10.1016/j.physrep.2016.06.005}{\emph{Phys. Rept.}
  {\bfseries 643} (2016) 1} [\href{https://arxiv.org/abs/1510.07633}{{\ttfamily
  1510.07633}}].

\bibitem{DiLuzio:2020wdo}
L.~Di~Luzio, M.~Giannotti, E.~Nardi and L.~Visinelli, \emph{{The landscape of
  QCD axion models}},
  \href{https://doi.org/10.1016/j.physrep.2020.06.002}{\emph{Phys. Rept.}
  {\bfseries 870} (2020) 1} [\href{https://arxiv.org/abs/2003.01100}{{\ttfamily
  2003.01100}}].

\bibitem{Muong-2:2021ojo}
{\scshape Muon g-2} collaboration, \emph{{Measurement of the Positive Muon
  Anomalous Magnetic Moment to 0.46 ppm}},
  \href{https://doi.org/10.1103/PhysRevLett.126.141801}{\emph{Phys. Rev. Lett.}
  {\bfseries 126} (2021) 141801}
  [\href{https://arxiv.org/abs/2104.03281}{{\ttfamily 2104.03281}}].

\bibitem{Muong-2:2006rrc}
{\scshape Muon g-2} collaboration, \emph{{Final Report of the Muon E821
  Anomalous Magnetic Moment Measurement at BNL}},
  \href{https://doi.org/10.1103/PhysRevD.73.072003}{\emph{Phys. Rev. D}
  {\bfseries 73} (2006) 072003}
  [\href{https://arxiv.org/abs/hep-ex/0602035}{{\ttfamily hep-ex/0602035}}].

\bibitem{Aoyama:2020ynm}
T.~Aoyama et~al., \emph{{The anomalous magnetic moment of the muon in the
  Standard Model}},
  \href{https://doi.org/10.1016/j.physrep.2020.07.006}{\emph{Phys. Rept.}
  {\bfseries 887} (2020) 1} [\href{https://arxiv.org/abs/2006.04822}{{\ttfamily
  2006.04822}}].

\bibitem{Aoyama:2012wk}
T.~Aoyama, M.~Hayakawa, T.~Kinoshita and M.~Nio, \emph{{Complete Tenth-Order
  QED Contribution to the Muon g-2}},
  \href{https://doi.org/10.1103/PhysRevLett.109.111808}{\emph{Phys. Rev. Lett.}
  {\bfseries 109} (2012) 111808}
  [\href{https://arxiv.org/abs/1205.5370}{{\ttfamily 1205.5370}}].

\bibitem{Aoyama:2019ryr}
T.~Aoyama, T.~Kinoshita and M.~Nio, \emph{{Theory of the Anomalous Magnetic
  Moment of the Electron}},
  \href{https://doi.org/10.3390/atoms7010028}{\emph{Atoms} {\bfseries 7} (2019)
  28}.

\bibitem{Czarnecki:2002nt}
A.~Czarnecki, W.J.~Marciano and A.~Vainshtein, \emph{{Refinements in
  electroweak contributions to the muon anomalous magnetic moment}},
  \href{https://doi.org/10.1103/PhysRevD.67.073006}{\emph{Phys. Rev. D}
  {\bfseries 67} (2003) 073006}
  [\href{https://arxiv.org/abs/hep-ph/0212229}{{\ttfamily hep-ph/0212229}}].

\bibitem{Gnendiger:2013pva}
C.~Gnendiger, D.~St\"ockinger and H.~St\"ockinger-Kim, \emph{{The electroweak
  contributions to $(g-2)_\mu$ after the Higgs boson mass measurement}},
  \href{https://doi.org/10.1103/PhysRevD.88.053005}{\emph{Phys. Rev. D}
  {\bfseries 88} (2013) 053005}
  [\href{https://arxiv.org/abs/1306.5546}{{\ttfamily 1306.5546}}].

\bibitem{Davier:2017zfy}
M.~Davier, A.~Hoecker, B.~Malaescu and Z.~Zhang, \emph{{Reevaluation of the
  hadronic vacuum polarisation contributions to the Standard Model predictions
  of the muon $g-2$ and ${\alpha (m_Z^2)}$ using newest hadronic cross-section
  data}}, \href{https://doi.org/10.1140/epjc/s10052-017-5161-6}{\emph{Eur.
  Phys. J. C} {\bfseries 77} (2017) 827}
  [\href{https://arxiv.org/abs/1706.09436}{{\ttfamily 1706.09436}}].

\bibitem{Keshavarzi:2018mgv}
A.~Keshavarzi, D.~Nomura and T.~Teubner, \emph{{Muon $g-2$ and $\alpha(M_Z^2)$:
  a new data-based analysis}},
  \href{https://doi.org/10.1103/PhysRevD.97.114025}{\emph{Phys. Rev. D}
  {\bfseries 97} (2018) 114025}
  [\href{https://arxiv.org/abs/1802.02995}{{\ttfamily 1802.02995}}].

\bibitem{Colangelo:2018mtw}
G.~Colangelo, M.~Hoferichter and P.~Stoffer, \emph{{Two-pion contribution to
  hadronic vacuum polarization}},
  \href{https://doi.org/10.1007/JHEP02(2019)006}{\emph{JHEP} {\bfseries 02}
  (2019) 006} [\href{https://arxiv.org/abs/1810.00007}{{\ttfamily
  1810.00007}}].

\bibitem{Hoferichter:2019mqg}
M.~Hoferichter, B.-L.~Hoid and B.~Kubis, \emph{{Three-pion contribution to
  hadronic vacuum polarization}},
  \href{https://doi.org/10.1007/JHEP08(2019)137}{\emph{JHEP} {\bfseries 08}
  (2019) 137} [\href{https://arxiv.org/abs/1907.01556}{{\ttfamily
  1907.01556}}].

\bibitem{Davier:2019can}
M.~Davier, A.~Hoecker, B.~Malaescu and Z.~Zhang, \emph{{A new evaluation of the
  hadronic vacuum polarisation contributions to the muon anomalous magnetic
  moment and to $\mathbf{\boldsymbol\alpha(m_Z^2)}$}},
  \href{https://doi.org/10.1140/epjc/s10052-020-7792-2}{\emph{Eur. Phys. J. C}
  {\bfseries 80} (2020) 241}
  [\href{https://arxiv.org/abs/1908.00921}{{\ttfamily 1908.00921}}].

\bibitem{Keshavarzi:2019abf}
A.~Keshavarzi, D.~Nomura and T.~Teubner, \emph{{$g-2$ of charged leptons,
  $\alpha (M^2_Z)$ , and the hyperfine splitting of muonium}},
  \href{https://doi.org/10.1103/PhysRevD.101.014029}{\emph{Phys. Rev. D}
  {\bfseries 101} (2020) 014029}
  [\href{https://arxiv.org/abs/1911.00367}{{\ttfamily 1911.00367}}].

\bibitem{Kurz:2014wya}
A.~Kurz, T.~Liu, P.~Marquard and M.~Steinhauser, \emph{{Hadronic contribution
  to the muon anomalous magnetic moment to next-to-next-to-leading order}},
  \href{https://doi.org/10.1016/j.physletb.2014.05.043}{\emph{Phys. Lett. B}
  {\bfseries 734} (2014) 144}
  [\href{https://arxiv.org/abs/1403.6400}{{\ttfamily 1403.6400}}].

\bibitem{Melnikov:2003xd}
K.~Melnikov and A.~Vainshtein, \emph{{Hadronic light-by-light scattering
  contribution to the muon anomalous magnetic moment revisited}},
  \href{https://doi.org/10.1103/PhysRevD.70.113006}{\emph{Phys. Rev. D}
  {\bfseries 70} (2004) 113006}
  [\href{https://arxiv.org/abs/hep-ph/0312226}{{\ttfamily hep-ph/0312226}}].

\bibitem{Masjuan:2017tvw}
P.~Masjuan and P.~Sanchez-Puertas, \emph{{Pseudoscalar-pole contribution to the
  $(g_{\mu}-2)$: a rational approach}},
  \href{https://doi.org/10.1103/PhysRevD.95.054026}{\emph{Phys. Rev. D}
  {\bfseries 95} (2017) 054026}
  [\href{https://arxiv.org/abs/1701.05829}{{\ttfamily 1701.05829}}].

\bibitem{Colangelo:2017fiz}
G.~Colangelo, M.~Hoferichter, M.~Procura and P.~Stoffer, \emph{{Dispersion
  relation for hadronic light-by-light scattering: two-pion contributions}},
  \href{https://doi.org/10.1007/JHEP04(2017)161}{\emph{JHEP} {\bfseries 04}
  (2017) 161} [\href{https://arxiv.org/abs/1702.07347}{{\ttfamily
  1702.07347}}].

\bibitem{Hoferichter:2018kwz}
M.~Hoferichter, B.-L.~Hoid, B.~Kubis, S.~Leupold and S.P.~Schneider,
  \emph{{Dispersion relation for hadronic light-by-light scattering: pion
  pole}}, \href{https://doi.org/10.1007/JHEP10(2018)141}{\emph{JHEP} {\bfseries
  10} (2018) 141} [\href{https://arxiv.org/abs/1808.04823}{{\ttfamily
  1808.04823}}].

\bibitem{Gerardin:2019vio}
A.~G\'erardin, H.B.~Meyer and A.~Nyffeler, \emph{{Lattice calculation of the
  pion transition form factor with $N_f=2+1$ Wilson quarks}},
  \href{https://doi.org/10.1103/PhysRevD.100.034520}{\emph{Phys. Rev. D}
  {\bfseries 100} (2019) 034520}
  [\href{https://arxiv.org/abs/1903.09471}{{\ttfamily 1903.09471}}].

\bibitem{Bijnens:2019ghy}
J.~Bijnens, N.~Hermansson-Truedsson and A.~Rodr\'\i{}guez-S\'anchez,
  \emph{{Short-distance constraints for the HLbL contribution to the muon
  anomalous magnetic moment}},
  \href{https://doi.org/10.1016/j.physletb.2019.134994}{\emph{Phys. Lett. B}
  {\bfseries 798} (2019) 134994}
  [\href{https://arxiv.org/abs/1908.03331}{{\ttfamily 1908.03331}}].

\bibitem{Colangelo:2019uex}
G.~Colangelo, F.~Hagelstein, M.~Hoferichter, L.~Laub and P.~Stoffer,
  \emph{{Longitudinal short-distance constraints for the hadronic
  light-by-light contribution to $(g-2)_\mu$ with large-$N_c$ Regge models}},
  \href{https://doi.org/10.1007/JHEP03(2020)101}{\emph{JHEP} {\bfseries 03}
  (2020) 101} [\href{https://arxiv.org/abs/1910.13432}{{\ttfamily
  1910.13432}}].

\bibitem{Blum:2019ugy}
T.~Blum, N.~Christ, M.~Hayakawa, T.~Izubuchi, L.~Jin, C.~Jung et~al.,
  \emph{{Hadronic Light-by-Light Scattering Contribution to the Muon Anomalous
  Magnetic Moment from Lattice QCD}},
  \href{https://doi.org/10.1103/PhysRevLett.124.132002}{\emph{Phys. Rev. Lett.}
  {\bfseries 124} (2020) 132002}
  [\href{https://arxiv.org/abs/1911.08123}{{\ttfamily 1911.08123}}].

\bibitem{Colangelo:2014qya}
G.~Colangelo, M.~Hoferichter, A.~Nyffeler, M.~Passera and P.~Stoffer,
  \emph{{Remarks on higher-order hadronic corrections to the muon
  g\ensuremath{-}2}},
  \href{https://doi.org/10.1016/j.physletb.2014.06.012}{\emph{Phys. Lett. B}
  {\bfseries 735} (2014) 90} [\href{https://arxiv.org/abs/1403.7512}{{\ttfamily
  1403.7512}}].

\bibitem{Parker:2018vye}
R.H.~Parker, C.~Yu, W.~Zhong, B.~Estey and H.~M\"uller, \emph{{Measurement of
  the fine-structure constant as a test of the Standard Model}},
  \href{https://doi.org/10.1126/science.aap7706}{\emph{Science} {\bfseries 360}
  (2018) 191} [\href{https://arxiv.org/abs/1812.04130}{{\ttfamily
  1812.04130}}].

\bibitem{Hanneke:2008tm}
D.~Hanneke, S.~Fogwell and G.~Gabrielse, \emph{{New Measurement of the Electron
  Magnetic Moment and the Fine Structure Constant}},
  \href{https://doi.org/10.1103/PhysRevLett.100.120801}{\emph{Phys. Rev. Lett.}
  {\bfseries 100} (2008) 120801}
  [\href{https://arxiv.org/abs/0801.1134}{{\ttfamily 0801.1134}}].

\bibitem{Hanneke:2010au}
D.~Hanneke, S.F.~Hoogerheide and G.~Gabrielse, \emph{{Cavity Control of a
  Single-Electron Quantum Cyclotron: Measuring the Electron Magnetic Moment}},
  \href{https://doi.org/10.1103/PhysRevA.83.052122}{\emph{Phys. Rev. A}
  {\bfseries 83} (2011) 052122}
  [\href{https://arxiv.org/abs/1009.4831}{{\ttfamily 1009.4831}}].

\bibitem{Morel:2020dww}
L.~Morel, Z.~Yao, P.~Clad\'e and S.~Guellati-Kh\'elifa, \emph{{Determination of
  the fine-structure constant with an accuracy of 81 parts per trillion}},
  \href{https://doi.org/10.1038/s41586-020-2964-7}{\emph{Nature} {\bfseries
  588} (2020) 61}.

\bibitem{CDF:2022hxs}
{\scshape CDF} collaboration, \emph{{High-precision measurement of the W boson
  mass with the CDF II detector}},
  \href{https://doi.org/10.1126/science.abk1781}{\emph{Science} {\bfseries 376}
  (2022) 170}.

\bibitem{ParticleDataGroup:2022pth}
{\scshape Particle Data Group} collaboration, \emph{{Review of Particle
  Physics}}, \href{https://doi.org/10.1093/ptep/ptac097}{\emph{PTEP} {\bfseries
  2022} (2022) 083C01}.

\bibitem{Athron:2022qpo}
P.~Athron, A.~Fowlie, C.-T.~Lu, L.~Wu, Y.~Wu and B.~Zhu, \emph{{The $W$ boson
  Mass and Muon $g-2$: Hadronic Uncertainties or New Physics?}},
  \href{https://arxiv.org/abs/2204.03996}{{\ttfamily 2204.03996}}.

\bibitem{Popov:2016fzr}
O.~Popov and G.A.~White, \emph{{One Leptoquark to unify them? Neutrino masses
  and unification in the light of $(g-2)_\mu$, $R_{D^{(\star)}}$ and $R_K$
  anomalies}},
  \href{https://doi.org/10.1016/j.nuclphysb.2017.08.007}{\emph{Nucl. Phys. B}
  {\bfseries 923} (2017) 324}
  [\href{https://arxiv.org/abs/1611.04566}{{\ttfamily 1611.04566}}].

\bibitem{DelleRose:2020oaa}
L.~Delle~Rose, S.~Khalil and S.~Moretti, \emph{{Explaining electron and muon
  $g-2$ anomalies in an Aligned 2-Higgs Doublet Model with right-handed
  neutrinos}},
  \href{https://doi.org/10.1016/j.physletb.2021.136216}{\emph{Phys. Lett. B}
  {\bfseries 816} (2021) 136216}
  [\href{https://arxiv.org/abs/2012.06911}{{\ttfamily 2012.06911}}].

\bibitem{Cacciapaglia:2022xih}
G.~Cacciapaglia and F.~Sannino, \emph{{The W boson mass weighs in on the
  non-standard Higgs}},
  \href{https://doi.org/10.1016/j.physletb.2022.137232}{\emph{Phys. Lett. B}
  {\bfseries 832} (2022) 137232}
  [\href{https://arxiv.org/abs/2204.04514}{{\ttfamily 2204.04514}}].

\bibitem{Lee:2022nqz}
H.M.~Lee and K.~Yamashita, \emph{{A model of vector-like leptons for the muon
  $g-2$ and the W boson mass}},
  \href{https://doi.org/10.1140/epjc/s10052-022-10635-z}{\emph{Eur. Phys. J. C}
  {\bfseries 82} (2022) 661}
  [\href{https://arxiv.org/abs/2204.05024}{{\ttfamily 2204.05024}}].

\bibitem{Babu:2022pdn}
K.S.~Babu, S.~Jana and V.P.~K., \emph{{Correlating W-Boson Mass Shift with Muon
  g-2 in the Two Higgs Doublet Model}},
  \href{https://doi.org/10.1103/PhysRevLett.129.121803}{\emph{Phys. Rev. Lett.}
  {\bfseries 129} (2022) 121803}
  [\href{https://arxiv.org/abs/2204.05303}{{\ttfamily 2204.05303}}].

\bibitem{Balkin:2022glu}
R.~Balkin, E.~Madge, T.~Menzo, G.~Perez, Y.~Soreq and J.~Zupan, \emph{{On the
  implications of positive W mass shift}},
  \href{https://doi.org/10.1007/JHEP05(2022)133}{\emph{JHEP} {\bfseries 05}
  (2022) 133} [\href{https://arxiv.org/abs/2204.05992}{{\ttfamily
  2204.05992}}].

\bibitem{Ahn:2022xax}
Y.H.~Ahn, S.K.~Kang and R.~Ramos, \emph{{Implications of New CDF-II $W$ Boson
  Mass on Two Higgs Doublet Model}},
  \href{https://doi.org/10.1103/PhysRevD.106.055038}{\emph{Phys. Rev. D}
  {\bfseries 106} (2022) 055038}
  [\href{https://arxiv.org/abs/2204.06485}{{\ttfamily 2204.06485}}].

\bibitem{Kawamura:2022uft}
J.~Kawamura, S.~Okawa and Y.~Omura, \emph{{W boson mass and muon g-2 in a
  lepton portal dark matter model}},
  \href{https://doi.org/10.1103/PhysRevD.106.015005}{\emph{Phys. Rev. D}
  {\bfseries 106} (2022) 015005}
  [\href{https://arxiv.org/abs/2204.07022}{{\ttfamily 2204.07022}}].

\bibitem{Ghoshal:2022vzo}
A.~Ghoshal, N.~Okada, S.~Okada, D.~Raut, Q.~Shafi and A.~Thapa, \emph{{Type III
  seesaw with R-parity violation in light of $m_W$ (CDF)}},
  \href{https://arxiv.org/abs/2204.07138}{{\ttfamily 2204.07138}}.

\bibitem{Kanemura:2022ahw}
S.~Kanemura and K.~Yagyu, \emph{{Implication of the W boson mass anomaly at CDF
  II in the Higgs triplet model with a mass difference}},
  \href{https://doi.org/10.1016/j.physletb.2022.137217}{\emph{Phys. Lett. B}
  {\bfseries 831} (2022) 137217}
  [\href{https://arxiv.org/abs/2204.07511}{{\ttfamily 2204.07511}}].

\bibitem{Chowdhury:2022moc}
T.A.~Chowdhury, J.~Heeck, A.~Thapa and S.~Saad, \emph{{W boson mass shift and
  muon magnetic moment in the Zee model}},
  \href{https://doi.org/10.1103/PhysRevD.106.035004}{\emph{Phys. Rev. D}
  {\bfseries 106} (2022) 035004}
  [\href{https://arxiv.org/abs/2204.08390}{{\ttfamily 2204.08390}}].

\bibitem{Borah:2022zim}
D.~Borah, S.~Mahapatra and N.~Sahu, \emph{{Singlet-doublet fermion origin of
  dark matter, neutrino mass and W-mass anomaly}},
  \href{https://doi.org/10.1016/j.physletb.2022.137196}{\emph{Phys. Lett. B}
  {\bfseries 831} (2022) 137196}
  [\href{https://arxiv.org/abs/2204.09671}{{\ttfamily 2204.09671}}].

\bibitem{Lee:2022gyf}
S.~Lee, K.~Cheung, J.~Kim, C.-T.~Lu and J.~Song, \emph{{Status of the
  two-Higgs-doublet model in light of the CDF mW measurement}},
  \href{https://doi.org/10.1103/PhysRevD.106.075013}{\emph{Phys. Rev. D}
  {\bfseries 106} (2022) 075013}
  [\href{https://arxiv.org/abs/2204.10338}{{\ttfamily 2204.10338}}].

\bibitem{Abouabid:2022lpg}
H.~Abouabid, A.~Arhrib, R.~Benbrik, M.~Krab and M.~Ouchemhou, \emph{{Is the new
  CDF $M_W$ measurement consistent with the two higgs doublet model?}},
  \href{https://arxiv.org/abs/2204.12018}{{\ttfamily 2204.12018}}.

\bibitem{Kim:2022hvh}
J.~Kim, S.~Lee, P.~Sanyal and J.~Song, \emph{{CDF W-boson mass and muon g-2 in
  a type-X two-Higgs-doublet model with a Higgs-phobic light pseudoscalar}},
  \href{https://doi.org/10.1103/PhysRevD.106.035002}{\emph{Phys. Rev. D}
  {\bfseries 106} (2022) 035002}
  [\href{https://arxiv.org/abs/2205.01701}{{\ttfamily 2205.01701}}].

\bibitem{Chowdhury:2022dps}
T.A.~Chowdhury and S.~Saad, \emph{{Leptoquark-vectorlike quark model for the
  CDF mW, (g-2)\ensuremath{\mu}, RK(*) anomalies, and neutrino masses}},
  \href{https://doi.org/10.1103/PhysRevD.106.055017}{\emph{Phys. Rev. D}
  {\bfseries 106} (2022) 055017}
  [\href{https://arxiv.org/abs/2205.03917}{{\ttfamily 2205.03917}}].

\bibitem{Hessenberger:2022tcx}
S.~Hessenberger, S.~Hessenberger and W.~Hollik, \emph{{Two-loop improved
  predictions for $\mathbf {M_W}$ and $\mathbf {sin^2\theta _{eff}}$ in
  Two-Higgs-Doublet models}},
  \href{https://doi.org/10.1140/epjc/s10052-022-10933-6}{\emph{Eur. Phys. J. C}
  {\bfseries 82} (2022) 970}
  [\href{https://arxiv.org/abs/2207.03845}{{\ttfamily 2207.03845}}].

\bibitem{Saez:2021qta}
B.D.~S\'aez and K.~Ghorbani, \emph{{Singlet scalars as dark matter and the muon
  ($g-2$) anomaly}},
  \href{https://doi.org/10.1016/j.physletb.2021.136750}{\emph{Phys. Lett. B}
  {\bfseries 823} (2021) 136750}
  [\href{https://arxiv.org/abs/2107.08945}{{\ttfamily 2107.08945}}].

\bibitem{Chakrabarty:2022gqi}
N.~Chakrabarty, I.~Chakraborty, D.K.~Ghosh and G.~Saha, \emph{{Muon $g-2$ and
  $W$-mass in a framework of colored scalars: an LHC perspective}},
  \href{https://arxiv.org/abs/2212.14458}{{\ttfamily 2212.14458}}.

\bibitem{Agrawal:2022wjm}
P.~Agrawal, D.E.~Kaplan, O.~Kim, S.~Rajendran and M.~Reig, \emph{{Searching for
  axion forces with precision precession in storage rings}},
  \href{https://arxiv.org/abs/2210.17547}{{\ttfamily 2210.17547}}.

\bibitem{Chakrabarty:2022voz}
N.~Chakrabarty, \emph{{The muon $g-2$ and $W$-mass anomalies explained and the
  electroweak vacuum stabilised by extending the minimal Type-II seesaw}},
  \href{https://arxiv.org/abs/2206.11771}{{\ttfamily 2206.11771}}.

\bibitem{Abdallah:2022shy}
W.~Abdallah, R.~Gandhi and S.~Roy, \emph{{LSND and MiniBooNE as guideposts to
  understanding the muon $g-2$ results and the CDF II $W$ mass measurement}},
  \href{https://arxiv.org/abs/2208.02264}{{\ttfamily 2208.02264}}.

\bibitem{Heckman:2022the}
J.J.~Heckman, \emph{{Extra W-boson mass from a D3-brane}},
  \href{https://doi.org/10.1016/j.physletb.2022.137387}{\emph{Phys. Lett. B}
  {\bfseries 833} (2022) 137387}
  [\href{https://arxiv.org/abs/2204.05302}{{\ttfamily 2204.05302}}].

\bibitem{Hiller:2019mou}
G.~Hiller, C.~Hormigos-Feliu, D.F.~Litim and T.~Steudtner, \emph{{Anomalous
  magnetic moments from asymptotic safety}},
  \href{https://doi.org/10.1103/PhysRevD.102.071901}{\emph{Phys. Rev. D}
  {\bfseries 102} (2020) 071901}
  [\href{https://arxiv.org/abs/1910.14062}{{\ttfamily 1910.14062}}].

\bibitem{Han:2018znu}
X.-F.~Han, T.~Li, L.~Wang and Y.~Zhang, \emph{{Simple interpretations of lepton
  anomalies in the lepton-specific inert two-Higgs-doublet model}},
  \href{https://doi.org/10.1103/PhysRevD.99.095034}{\emph{Phys. Rev. D}
  {\bfseries 99} (2019) 095034}
  [\href{https://arxiv.org/abs/1812.02449}{{\ttfamily 1812.02449}}].

\bibitem{Zhitnitsky:1980tq}
A.R.~Zhitnitsky, \emph{{On Possible Suppression of the Axion Hadron
  Interactions. (In Russian)}}, {\emph{Sov. J. Nucl. Phys.} {\bfseries 31}
  (1980) 260}.

\bibitem{Dine:1981rt}
M.~Dine, W.~Fischler and M.~Srednicki, \emph{{A Simple Solution to the Strong
  CP Problem with a Harmless Axion}},
  \href{https://doi.org/10.1016/0370-2693(81)90590-6}{\emph{Phys. Lett. B}
  {\bfseries 104} (1981) 199}.

\bibitem{ParticleDataGroup:2014cgo}
{\scshape Particle Data Group} collaboration, \emph{{Review of Particle
  Physics}}, \href{https://doi.org/10.1088/1674-1137/38/9/090001}{\emph{Chin.
  Phys. C} {\bfseries 38} (2014) 090001}.

\bibitem{CMS:2018qvj}
{\scshape CMS} collaboration, \emph{{Search for an exotic decay of the Higgs
  boson to a pair of light pseudoscalars in the final state of two muons and
  two $\tau$ leptons in proton-proton collisions at $ \sqrt{s}=13 $ TeV}},
  \href{https://doi.org/10.1007/JHEP11(2018)018}{\emph{JHEP} {\bfseries 11}
  (2018) 018} [\href{https://arxiv.org/abs/1805.04865}{{\ttfamily
  1805.04865}}].

\bibitem{Raffelt:2006cw}
G.G.~Raffelt, \emph{{Astrophysical axion bounds}},
  \href{https://doi.org/10.1007/978-3-540-73518-2_3}{\emph{Lect. Notes Phys.}
  {\bfseries 741} (2008) 51}
  [\href{https://arxiv.org/abs/hep-ph/0611350}{{\ttfamily hep-ph/0611350}}].

\bibitem{Arvanitaki:2009fg}
A.~Arvanitaki, S.~Dimopoulos, S.~Dubovsky, N.~Kaloper and J.~March-Russell,
  \emph{{String Axiverse}},
  \href{https://doi.org/10.1103/PhysRevD.81.123530}{\emph{Phys. Rev. D}
  {\bfseries 81} (2010) 123530}
  [\href{https://arxiv.org/abs/0905.4720}{{\ttfamily 0905.4720}}].

\bibitem{Kim:1979if}
J.E.~Kim, \emph{{Weak Interaction Singlet and Strong CP Invariance}},
  \href{https://doi.org/10.1103/PhysRevLett.43.103}{\emph{Phys. Rev. Lett.}
  {\bfseries 43} (1979) 103}.

\bibitem{Shifman:1979if}
M.A.~Shifman, A.I.~Vainshtein and V.I.~Zakharov, \emph{{Can Confinement Ensure
  Natural CP Invariance of Strong Interactions?}},
  \href{https://doi.org/10.1016/0550-3213(80)90209-6}{\emph{Nucl. Phys. B}
  {\bfseries 166} (1980) 493}.

\bibitem{Ahmadvand:2021vxs}
M.~Ahmadvand, \emph{{Filtered asymmetric dark matter during the Peccei-Quinn
  phase transition}},
  \href{https://doi.org/10.1007/JHEP10(2021)109}{\emph{JHEP} {\bfseries 10}
  (2021) 109} [\href{https://arxiv.org/abs/2108.00958}{{\ttfamily
  2108.00958}}].

\bibitem{Espriu:2015mfa}
D.~Espriu, F.~Mescia and A.~Renau, \emph{{Axion-Higgs interplay in the two
  Higgs-doublet model}},
  \href{https://doi.org/10.1103/PhysRevD.92.095013}{\emph{Phys. Rev. D}
  {\bfseries 92} (2015) 095013}
  [\href{https://arxiv.org/abs/1503.02953}{{\ttfamily 1503.02953}}].

\bibitem{Branco:2011iw}
G.C.~Branco, P.M.~Ferreira, L.~Lavoura, M.N.~Rebelo, M.~Sher and J.P.~Silva,
  \emph{{Theory and phenomenology of two-Higgs-doublet models}},
  \href{https://doi.org/10.1016/j.physrep.2012.02.002}{\emph{Phys. Rept.}
  {\bfseries 516} (2012) 1} [\href{https://arxiv.org/abs/1106.0034}{{\ttfamily
  1106.0034}}].

\bibitem{MSSMWorkingGroup:1998fiq}
{\scshape MSSM Working Group} collaboration, \emph{{The Minimal supersymmetric
  standard model: Group summary report}},  in \emph{{GDR (Groupement De
  Recherche) - Supersymetrie}}, 12, 1998
  [\href{https://arxiv.org/abs/hep-ph/9901246}{{\ttfamily hep-ph/9901246}}].

\bibitem{ALEPH:2013htx}
{\scshape ALEPH, DELPHI, L3, OPAL, LEP} collaboration, \emph{{Search for
  Charged Higgs bosons: Combined Results Using LEP Data}},
  \href{https://doi.org/10.1140/epjc/s10052-013-2463-1}{\emph{Eur. Phys. J. C}
  {\bfseries 73} (2013) 2463}
  [\href{https://arxiv.org/abs/1301.6065}{{\ttfamily 1301.6065}}].

\bibitem{CDF:2009efz}
{\scshape CDF} collaboration, \emph{{Search for charged Higgs bosons in decays
  of top quarks in p anti-p collisions at s**(1/2) = 1.96 TeV}},
  \href{https://doi.org/10.1103/PhysRevLett.103.101803}{\emph{Phys. Rev. Lett.}
  {\bfseries 103} (2009) 101803}
  [\href{https://arxiv.org/abs/0907.1269}{{\ttfamily 0907.1269}}].

\bibitem{LHCHiggsCrossSectionWorkingGroup:2016ypw}
{\scshape LHC Higgs Cross Section Working Group} collaboration, \emph{{Handbook
  of LHC Higgs Cross Sections: 4. Deciphering the Nature of the Higgs Sector}},
   \href{https://arxiv.org/abs/1610.07922}{{\ttfamily 1610.07922}}.

\bibitem{Ahmed:2016otz}
T.~Ahmed, M.~Bonvini, M.C.~Kumar, P.~Mathews, N.~Rana, V.~Ravindran et~al.,
  \emph{{Pseudo-scalar Higgs boson production at N$^3$ LO$_{\text {A}}$ +N$^3$
  LL $'$}}, \href{https://doi.org/10.1140/epjc/s10052-016-4510-1}{\emph{Eur.
  Phys. J. C} {\bfseries 76} (2016) 663}
  [\href{https://arxiv.org/abs/1606.00837}{{\ttfamily 1606.00837}}].

\bibitem{Peskin:1990zt}
M.E.~Peskin and T.~Takeuchi, \emph{{A New constraint on a strongly interacting
  Higgs sector}}, \href{https://doi.org/10.1103/PhysRevLett.65.964}{\emph{Phys.
  Rev. Lett.} {\bfseries 65} (1990) 964}.

\bibitem{Peskin:1991sw}
M.E.~Peskin and T.~Takeuchi, \emph{{Estimation of oblique electroweak
  corrections}}, \href{https://doi.org/10.1103/PhysRevD.46.381}{\emph{Phys.
  Rev. D} {\bfseries 46} (1992) 381}.

\bibitem{Ghorbani:2022vtv}
K.~Ghorbani and P.~Ghorbani, \emph{{W-boson mass anomaly from scale invariant
  2HDM}}, \href{https://doi.org/10.1016/j.nuclphysb.2022.115980}{\emph{Nucl.
  Phys. B} {\bfseries 984} (2022) 115980}
  [\href{https://arxiv.org/abs/2204.09001}{{\ttfamily 2204.09001}}].

\bibitem{Foot:2013hna}
R.~Foot, A.~Kobakhidze, K.L.~McDonald and R.R.~Volkas, \emph{{Poincar\'e
  protection for a natural electroweak scale}},
  \href{https://doi.org/10.1103/PhysRevD.89.115018}{\emph{Phys. Rev. D}
  {\bfseries 89} (2014) 115018}
  [\href{https://arxiv.org/abs/1310.0223}{{\ttfamily 1310.0223}}].

\bibitem{Patel:2015tea}
H.H.~Patel, \emph{{Package-X: A Mathematica package for the analytic
  calculation of one-loop integrals}},
  \href{https://doi.org/10.1016/j.cpc.2015.08.017}{\emph{Comput. Phys. Commun.}
  {\bfseries 197} (2015) 276}
  [\href{https://arxiv.org/abs/1503.01469}{{\ttfamily 1503.01469}}].

\bibitem{Keshavarzi:2021eqa}
A.~Keshavarzi, K.S.~Khaw and T.~Yoshioka, \emph{{Muon $g-2$: A review}},
  \href{https://doi.org/10.1016/j.nuclphysb.2022.115675}{\emph{Nucl. Phys. B}
  {\bfseries 975} (2022) 115675}
  [\href{https://arxiv.org/abs/2106.06723}{{\ttfamily 2106.06723}}].

\bibitem{Barr:1990vd}
S.M.~Barr and A.~Zee, \emph{{Electric Dipole Moment of the Electron and of the
  Neutron}}, \href{https://doi.org/10.1103/PhysRevLett.65.21}{\emph{Phys. Rev.
  Lett.} {\bfseries 65} (1990) 21}.

\bibitem{Ilisie:2015tra}
V.~Ilisie, \emph{{New Barr-Zee contributions to $\mathbf{(g-2)_\mu}$ in
  two-Higgs-doublet models}},
  \href{https://doi.org/10.1007/JHEP04(2015)077}{\emph{JHEP} {\bfseries 04}
  (2015) 077} [\href{https://arxiv.org/abs/1502.04199}{{\ttfamily
  1502.04199}}].

\bibitem{Cherchiglia:2016eui}
A.~Cherchiglia, P.~Kneschke, D.~St\"ockinger and H.~St\"ockinger-Kim,
  \emph{{The muon magnetic moment in the 2HDM: complete two-loop result}},
  \href{https://doi.org/10.1007/JHEP10(2021)242}{\emph{JHEP} {\bfseries 01}
  (2017) 007} [\href{https://arxiv.org/abs/1607.06292}{{\ttfamily
  1607.06292}}].

\bibitem{Broggio:2014mna}
A.~Broggio, E.J.~Chun, M.~Passera, K.M.~Patel and S.K.~Vempati, \emph{{Limiting
  two-Higgs-doublet models}},
  \href{https://doi.org/10.1007/JHEP11(2014)058}{\emph{JHEP} {\bfseries 11}
  (2014) 058} [\href{https://arxiv.org/abs/1409.3199}{{\ttfamily 1409.3199}}].

\bibitem{Jueid:2021avn}
A.~Jueid, J.~Kim, S.~Lee and J.~Song, \emph{{Type-X two-Higgs-doublet model in
  light of the muon g-2: Confronting Higgs boson and collider data}},
  \href{https://doi.org/10.1103/PhysRevD.104.095008}{\emph{Phys. Rev. D}
  {\bfseries 104} (2021) 095008}
  [\href{https://arxiv.org/abs/2104.10175}{{\ttfamily 2104.10175}}].

\bibitem{Botella:2022rte}
F.J.~Botella, F.~Cornet-Gomez, C.~Mir\'o and M.~Nebot, \emph{{Muon and electron
  $g-2$ anomalies in a flavor conserving 2HDM with an oblique view on the CDF
  $M_W$ value}},
  \href{https://doi.org/10.1140/epjc/s10052-022-10893-x}{\emph{Eur. Phys. J. C}
  {\bfseries 82} (2022) 915}
  [\href{https://arxiv.org/abs/2205.01115}{{\ttfamily 2205.01115}}].

\bibitem{MEG:2016leq}
{\scshape MEG} collaboration, \emph{{Search for the lepton flavour violating
  decay $\mu ^+ \rightarrow \mathrm {e}^+ \gamma $ with the full dataset of the
  MEG experiment}},
  \href{https://doi.org/10.1140/epjc/s10052-016-4271-x}{\emph{Eur. Phys. J. C}
  {\bfseries 76} (2016) 434}
  [\href{https://arxiv.org/abs/1605.05081}{{\ttfamily 1605.05081}}].

\end{thebibliography}\endgroup

\end{document}